\documentclass[useAMS,usenatbib]{mnras}

\usepackage{graphicx}\usepackage{epsfig}\usepackage{epsf}
\usepackage{amsmath}\usepackage{amssymb}\usepackage{stfloats}
\usepackage{soul}

\title[SN~2017hcc]{High-resolution spectroscopy of SN~2017hcc and its
  blueshifted line profiles from post-shock dust formation}

\author[Smith et al.]{Nathan Smith$^{1}$\thanks{E-mail:
    nathans@as.arizona.edu} and Jennifer E. Andrews  \\
  $^{1}$Steward Observatory, University of Arizona, 933 N. Cherry
  Ave., Tucson, AZ 85721, USA}

\begin{document}

\pagerange{\pageref{firstpage}--\pageref{lastpage}} \pubyear{2012}
\maketitle
\label{firstpage}

\begin{abstract}
  SN~2017hcc was remarkable for being a nearby and strongly polarized
  superluminous Type~IIn supernova (SN).  We obtained high-resolution echelle spectra that we combine with other spectra
  to investigate its line profile evolution. All epochs reveal narrow
  P~Cygni components from pre-shock circumstellar material (CSM),
  indicating an axisymmetric outflow from the progenitor of 40-50 km
  s$^{-1}$.  Intermediate-width and broad components exhibit the
  classic evolution seen in luminous SNe~IIn: symmetric Lorentzian
  profiles from pre-shock CSM lines broadened by
  electron scattering at early times, transitioning at late times to multi-component,
  irregular profiles coming from the SN ejecta and post-shock
  shell.  As in many SNe~IIn, profiles show a progressively increasing
  blueshift, with a clear flux deficit in red wings of the
  intermediate and broad velocity components after day 200.  This
  blueshift develops after the continuum luminosity fades, and in the
  intermediate-width component, persists at late times even after the
  SN ejecta fade.  In SN~2017hcc, the blueshift cannot be explained as occultation 
  by the SN photosphere, pre-shock acceleration of CSM, or a lopsided
  explosion or CSM.  Instead, the blueshift arises from dust
  formation in the post-shock shell and in the SN ejecta.  The effect
  has a wavelength dependence characteristic of dust, exhibiting an
  extinction law consistent with large grains.  Thus, SN~2017hcc
  experienced post-shock dust formation and had a mildly bipolar
  CSM shell, similar to SN~2010jl.   Like other
  superluminous SNe~IIn, the progenitor lost around 10 $M_{\odot}$ due to extreme
  eruptive mass loss in the decade before exploding.
\end{abstract}


\section{INTRODUCTION}

\subsection{Eruptive pre-SN mass loss}

Throughout their lives, massive stars shed mass through steady winds,
but these winds can be punctuated by episodic mass loss events
\citep{smith14}.  Observationally-derived wind mass-loss rates have
been revised downward compared to rates still used in most
stellar evolution codes, undermining some basic
predictions of single-star models
\citep{smith14}. Mass-loss rates have been reduced for
both line-driven winds of hot stars
\citep{bouret05,fullerton06,puls08,sundqvist19}, and more recently
also for dusty red supergiant (RSG) winds \citep{beasor20}.

On the other hand, there has been a shift to stronger influence of
eruptive mass loss \citep{so06} and binary mass exchange and
stripping \citep{sana12,moe17,gotberg18}.  Eruptive mass loss like
luminous blue variables (LBVs) or violent binary interaction are less
sensitive to metallicty than line-driven winds \citep{so06,gotberg17},
making them more relevant to supernovae (SNe) that
arise in lower-metallicity regions.  Eruptive
LBV-like mass loss and violent binary interaction may be related, since mergers and mass gainers are
needed to explain various properties of LBVs
\citep{justham14,st15,mojgan17,smith18}.

Eruptive mass loss has become especially prominent in the case of
strongly interacting SNe of Types~IIn (SNe~IIn) and Ibn (SNe~Ibn),
where observed signatures of interaction with dense circumstellar
material (CSM) indicate extreme and short-lived mass-loss
phases immediately preceding core collapse; see \citet{smith17} and
\citet{smith14} for reviews.  The specific mechanism(s) of the
eruptive pre-SN mass loss remains uncertain, but it requires a trigger
synchronized with core-collapse.  Some proposed mechanisms include
energy deposited in the envelope by wave driving in advanced nuclear
burning phases \citep{qs12,sq14,fuller17,fr18}, the pulsational pair
instability or other late-phase burning instabilities
\citep{woosley07,woosley17,sa14,renzo20}, or an inflation of the
progenitor's radius (perhaps caused by the previous mechanisms) that
triggers violent binary interaction like collisions or mergers before
core collapse \citep{sa14}.  Of these, only the last predicts highly
asymmetric distributions of CSM (disk-like or bipolar), relevant to
asymmetric line profile shapes and high polarization seen in SNe~IIn.

An extreme case of eruptive pre-SN mass loss is found in
super-luminous SNe (SLSNe) with exceptionally strong CSM interaction,
where the very high luminosity is thought to result from an explosion
with normal to high explosion energy (1-5$\times$10$^{51}$ erg) running to a
very high mass of CSM, usually of order a few to 20 M$_{\odot}$.  Some
well studied examples include SN~2006gy, SN~2006tf, SN~2008am,
SN~2003ma, SN~2015da, and SN~2010jl
\citep{smith07,smith10,sm07,manos11,rest11,Tartaglia2020,gall14,fransson14,jencson16,woosley07}
These events have high efficiency in converting ejecta kinetic energy
into luminosity of order $10\%-50\%$ or more \citep{vanmarle10}, with
total radiated energy budgets $\ga$10$^{51}$ erg.

\subsection{Dust formation in interacting SNe}

An important feature of strongly interacting SNe is the formation of a
cold dense shell (CDS) between the forward and reverse shocks
\citep{chugai01,chugai04,smith08tf}.  Efficient radiative cooling
causes the dense post-shock gas to collapse into a thin, dense, clumpy,
and probably well-mixed layer \citep{vanmarle10}.  This rapid cooling
that forms a dense shell is a unique feature of interacting SNe,
causing their defining narrow-line spectra and their high
luminosity.  This rapid cooling and high density may also
trigger early dust formation.

There are 3 common observational signatures of new dust
formation\footnote{Here ``dust formation'' may also mean pre-existing
  grains in the CSM that were incompletely destroyed by the forward
  shock and  that then regrow.} in SNe: (1) a strengthening IR excess
consistent with dust emission, (2) an increased rate of fading in the
optical continuum that may be attributed to increased extinction from
new dust, and (3) a progressive and systematic blueshift in
emission-line profiles caused by dust that preferentially obscures the
redshifted portions of the explosion.  Each of these alone is somewhat
ambiguous --- an IR excess could be due to an IR echo from CSM dust
\citep{fox11,andrews11}, the rate of fading can depend on other
factors unrelated to dust, and blueshifted line profiles can arise
from asymmetry in the ejecta or CSM, or optical depth effects at early
times --- but seeing all three together, as in the case of SN~1987A
\citep{danziger89,lucy89,gn89,wooden93,colgan94}, gives a strong
indication that new dust has formed.  Of these three, the third is
unique in helping to diagnose the location of the dust because of the
different charactistic expansion speeds in the unshocked CSM, SN
ejecta, and the post-shock CDS.  The characteristic
blueshift of emission lines is fairly common in interacting SNe, and
has been tied to dust formation in the SN ejecta
and post-shock CDS.

The first clear case of an interacting SN that showed all three
signatures of dust formation was SN~2006jc, where the line-profile
evolution of He~{\sc i} lines (this was a Type Ibn) indicated that new
dust formed in the post-shock CDS \citep{smith08jc}.  This blueshift
coincided in time with an IR excess and fading in the continuum, but
also a burst of X-ray emission and He~{\sc ii} $\lambda$4686 emission
\citep{immler08,smith08jc,dicarlo08}, indicating that the SN shock
crashing into a dense shell is what triggered the rapid onset of dust
formation only 50-100 days after discovery.  The rapid post-shock dust
formation may be analogous to what happens in dust-forming
colliding-wind binaries like the WC+OB system WR~140
\citep{hackwell79,williams90,monnier02} and $\eta$~Car
\citep{smith10eta}.  Similar blueshifted profiles have been seen in a
number of H-rich interacting SNe, including SN~2006tf
\citep{smith08tf}, SN~2005ip \citep{smith09,fox09,smith17ip}, SN~2007rt
\citep{trundle09}, SN~2007od \citep{andrews10}, SN~2010bt
\citep{eliasrosa18}, and SN~2010jl (discussed below).

The line profile blueshift seen in interacting SNe has usually been
interpreted as post-shock or ejecta dust formation.  Line profile
blueshift gives a more direct probe of the dust location, as
noted above, but it is harder to infer the dust mass from extinction's
effect on line profiles without detailed models and assumptions
\citep{bevan18,bevan20}.  One should realize that these options for the
location of the dust are not mutually exclusive.  Dust may form in
both the CDS and the inner SN ejecta at various times, and there may
also be pre-existing CSM dust.  In fact, we expect pre-existing dust
in the CSM for SNe~IIn because of extreme eruptive mass loss (like
LBVs or extreme RSGs), and those are the same progenitors most likely
to have the high CSM density to trigger efficient radiative cooling
and collapse of the CDS, which in turn permits efficient dust formation in the
post-shock region.  There has, however, been discussion in the
literature about alternatives to dust formation to explain the
blueshifts, as illustrated by the case of SN~2010jl.

\subsection{SN~2010jl}

SN~2010jl was among the nearest SLSNe~IIn, leading to considerable
observational attention with high-quality optical/IR photometry and
spectroscopy
\citep{andrews11,stoll11,smith11,smith12,zhang12,fox13,maeda13,gall14,fransson14,borish15,williams15,jencson16,sarangi18,chugai18,bevan20}.
While pre-existing CSM dust probably contributed to the observed IR
excess \citep{andrews11}, even early spectra revealed a
prominent blueshift in line profiles that strengthened with
time \citep{smith12,gall14}.

The systematic blueshift of line profiles
and their wavelength dependence (more pronouned at shorter
wavelengths) led to the suggestion that, like SN~2006jc, SN~2010jl
experienced new dust formation in the post-shock region of the CDS
\citep{smith12}.  Several additional studies confirmed these signs of
dust formation in the post-shock CDS and investigated the dust
properties, including additional late-time spectra, IR data, and
models \citep{gall14,sarangi18,chugai18,bevan20}.  Dust may have been 
pre-existing in the CSM, and dust may have
formed in the ejecta \citep{andrews11,bevan20}.

The conclusion that the blueshift in line profiles was
influenced by dust formation in the CDS was not, however, adopted
by all authors.  Namely, \citet{fransson14} proposed a different picture where
the blueshifted profiles were caused by acceleration of asymmetric pre-shock
CSM along the line of sight, and where the broad components were
caused by electron scattering of the narrow CSM emission from that
accelerated CSM.  This explanation did not account for why the narrow
components from the unshocked CSM were not blueshifted even though the
broader components had a blueshifted centroid, or why there was a
wavelength dependence to the asymmetry.  Subsequent radiative transfer
models of SN~2010jl's spectrum \citep{dessart15} found that
accelerated pre-shock CSM could not explain the observed blueshift in
line profile shape, or their behavior with time.  \citet{dessart15}
showed that while the early symmetric line profiles were caused
by electron scattering of narrow CSM emission, the broad blueshifted
profiles at later times arises in the post-shock CDS.  These
models revealed a blueshifted emission bump that could arise even
without dust, at least initially, because the
photosphere in the CSM interaction region could block the redshifted
side of the CDS, as noted earlier by \citet{smith12}.  Again, however,
this mechanism would not account for the observed wavelength
dependence of the blueshift \citep{smith12,gall14}.

If the systematic blueshift of the broader components were due mostly
to an optical depth effect in the CDS, as seen in these models for
SN~2010jl spectra up until about day 200 \citep{dessart15}, then there
is a clear prediction for the late-time evolution.  Namely, at late
times these lines should become more symmetric, as the optical depth
drops and the continuum luminosity fades, revealing emission from the
receding side \citep{smith12,dessart15}.  This did not happen in
SN~2010jl.  The continuum luminosity dropped around day 300, but the
blueshift persisted even in spectra beyond day 1000
\citep{fransson14}.  This would seem to clearly rule out high
continuum optical depths and electron scattering of accelerated CSM as
the explanation for the persistent blueshift, instead favoring dust formation in
the post-shock shell of SN~2010jl
\citep{smith12,gall14,sarangi18,bevan20}.

\begin{figure}
\includegraphics[width=3.0in]{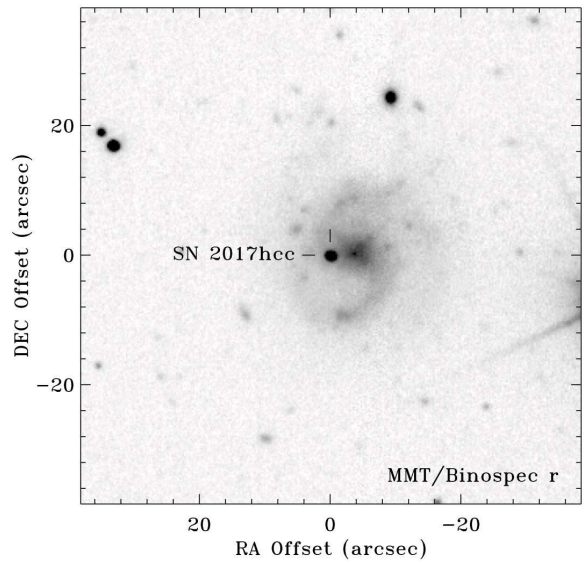}
\caption{A late-time $r$-band image of SN~2017hcc and its host galaxy
  taken on 2019 July 9 using Binospec on the MMT.  SN~2017hcc is found
  about 0.5 arcsec south and 4.5 arcsec east of the center of its
  spiral host galaxy.}
\label{fig:img}
\end{figure}

\subsection{SN~2017hcc}

SN~2017hcc was discovered on 2017 October 2 by the Asteroid
Terrestrial-impact Last Alert System (ATLAS; \citealt{tonry11}), and
was classified as a young Type~IIn event \citep{prieto17}. It was
reportedly located a few arcsec southeast of the center of an
anonymous host galaxy.  Figure~\ref{fig:img} shows an MMT/Binospec
image of the SN (see details below), indicating that it is located 0.5
arcsec south and 4.5 arcsec east of the center of its host
galaxy, and appears coincident with the leading edge of a spiral
arm. The SN was discovered within a few days of explosion, judging by
a non-detection a few days earlier.  Following \citet{prieto17}, who
presented early photometry, we adopt an explosion date of 2017 Oct
1.4, which we set as $t$ = 0 in this study.  As in \citet{prieto17},
we adopt a redshift of $z$=0.0168 and a distance of 73 Mpc.  We also
adopt $E(B-V)$=0.0285 mag for our line-of-sight through the Milky Way
interstellar medium \citep{schlafly11} to deredden our spectra,
although this has little impact on our analysis of line
profiles.  At this distance, its peak $V$ magnitude of 13.7 mag
implies an absolute magnitude of $-$20.7 mag \citep{prieto17}, making
SN~2017hcc a superluminous SN~IIn.  SN~2017hcc reached the peak of its $V$-band luminosity
about 40-45 days after explosion \citep{prieto17}.  This is a
relatively slow rise, although not as slow as the unusual
case of SN~2006gy, which took $\sim$70 days \citep{smith07}.
SN~2017hcc was not detected at early times in either X-rays with {\it
  Chandra} or in the radio \citep{chandra17,nc17}, although such
indicators are often faint at early times when X-ray and radio
emission are absorbed.

Perhaps the most notable property of SN~2017hcc so far is that
spectropolarimetry obtained soon after discovery by \citet{mauerhan17}
indicated very high continuum polarization of 4.8\% or more.  This may
be the highest level of polarization seen in any SN to date, and
points to significant asymmetry in the interaction region of this
SN~IIn.  This asymmetry indicated by polarization is relevant to the
interpretation of the observed evolution of emission-line profile
shapes for SN~2017hcc that we discuss below.  In section 2 we present
our new spectroscopic observations, in section 3 we present results
from the analysis of line profiles in our spectra, and in section 4 we
discuss the interpretation of these line profiles and corresponding implications for
the nature of the CSM and for dust formation in
SN~2017hcc.

%%%%%%%%%%%%%%%%%%%%%%%%%%%%%%%%%%%%%%%%%%%%%%%%%%%%%%%%%%%%%%%%%%%%%%%

\begin{table}\begin{center}\begin{minipage}{6.3in}%
    \caption{Low resolution spectroscopy of SN~2017hcc}\label{tab:spectab1}\scriptsize
\begin{tabular}{@{}lclcc}\hline\hline
UT Date      &day$^a$ &Tel./Instr.  &grating &$\Delta\lambda$ ($\mu$m) \\ \hline
%%%
%
2017 10 26   &25      &MMT/Blue     &1200    &0.57-0.70  \\
2017 10 27   &26      &Bok/BC       &300     &0.44-0.87  \\
2017 11 19   &49      &MMT/Blue     &1200    &0.57-0.70  \\
2017 12 09   &69      &Bok/BC       &300     &0.44-0.87  \\
2017 12 15   &75      &MMT/Blue     &1200    &0.57-0.70  \\
2017 12 28   &88      &MMT/Red      &1200    &0.62-0.70  \\
2018 01 01   &92      &Bok/BC       &300     &0.44-0.87  \\
2018 06 30   &271     &MMT/Blue     &1200    &0.57-0.70  \\
2018 08 31   &336     &MMT/MMIRS    &zJ/HK   &1.0-2.3  \\
2018 09 05   &340     &MMT/Blue     &300     &0.57-0.70  \\
2018 10 18   &382     &MMT/Bino     &600     &0.51-0.75  \\
2018 11 17   &412     &MMT/MMIRS    &zJ/HK   &1.0-2.3  \\
2019 07 08   &645     &MMT/Bino     &600     &0.51-0.75  \\
2019 11 02   &762     &MMT/Bino     &600     &0.51-0.75  \\
2020 01 27   &848     &Mag/IMACS    &1200    &0.63-0.67  \\
\hline
\end{tabular}
%%%%%%%%%%%%%%%%%%%%%%%%%%%%%%%%%%%%%%%%%%%%%%%%%%%%%%%%%%%%%%%%%%%%%%%%%
%%%%%%%%%%%%%%%%%%%%%%%%%%%%
\medskip
%%%%%%%%%%%%%%%%%%%%%%%%%%%%%%%%%%%%%%%%%%%%%%%%%%%%%%%%%%%%%%%%%%%%%%%%
\end{minipage}\end{center}
$^a$We adopt 2017 Oct 1 as day zero, close to the explosion date inferred by \citet{prieto17}.
\end{table}

\section{OBSERVATIONS}

\subsection{Optical image}

We obtained a deep $r$-band image of SN~2017hcc on 2019 July 9 (day
646) using the imaging mode of Binospec \citep{fabricant19} on the Multiple Mirror
Telescope (MMT). The image was reduced using a custom python pipeline\footnote{{\tt https://github.com/KerryPaterson/}} and is shown in Figure~\ref{fig:img}.
SN~2017hcc is located about 0.5 arcsec south and 4.5 arcsec east of
the center of its spiral host galaxy (a foreground star or cluster
projected near the galaxy nucleus would shift the centroid of the
galaxy light slightly to the east in lower-resolution images).
SN~2017hcc is therefore projected a total of 4.53 arcsec from center,
or $\sim$1.6 kpc from the nucleus at the adopted distance of 73 Mpc.
Although the host galaxy is anonymous and does not appear in the NASA
Extragalactic Database, according to SIMBAD\footnote{{\tt
    http://simbad.u-strasbg.fr/simbad/}}, it is coincident with the UV
source GALEX 2674128878581058535.  The coordinates listed for this
GALEX source are offset a few arcsec from the apparent center of the
galaxy.  SN~2017hcc appears to be located on the leading edge of an
inner spiral arm or bar, although the ground-based angular resolution
is insufficient to determine its proximity to any dust lanes or star
forming regions.
%We also obtained one deep image with comparable seeing using IMACS on
%the Magellan Clay Telescope, but this image is not shown because it
%contained a bad ghost image near the host galaxy.

\begin{figure*}
\includegraphics[width=6.0in]{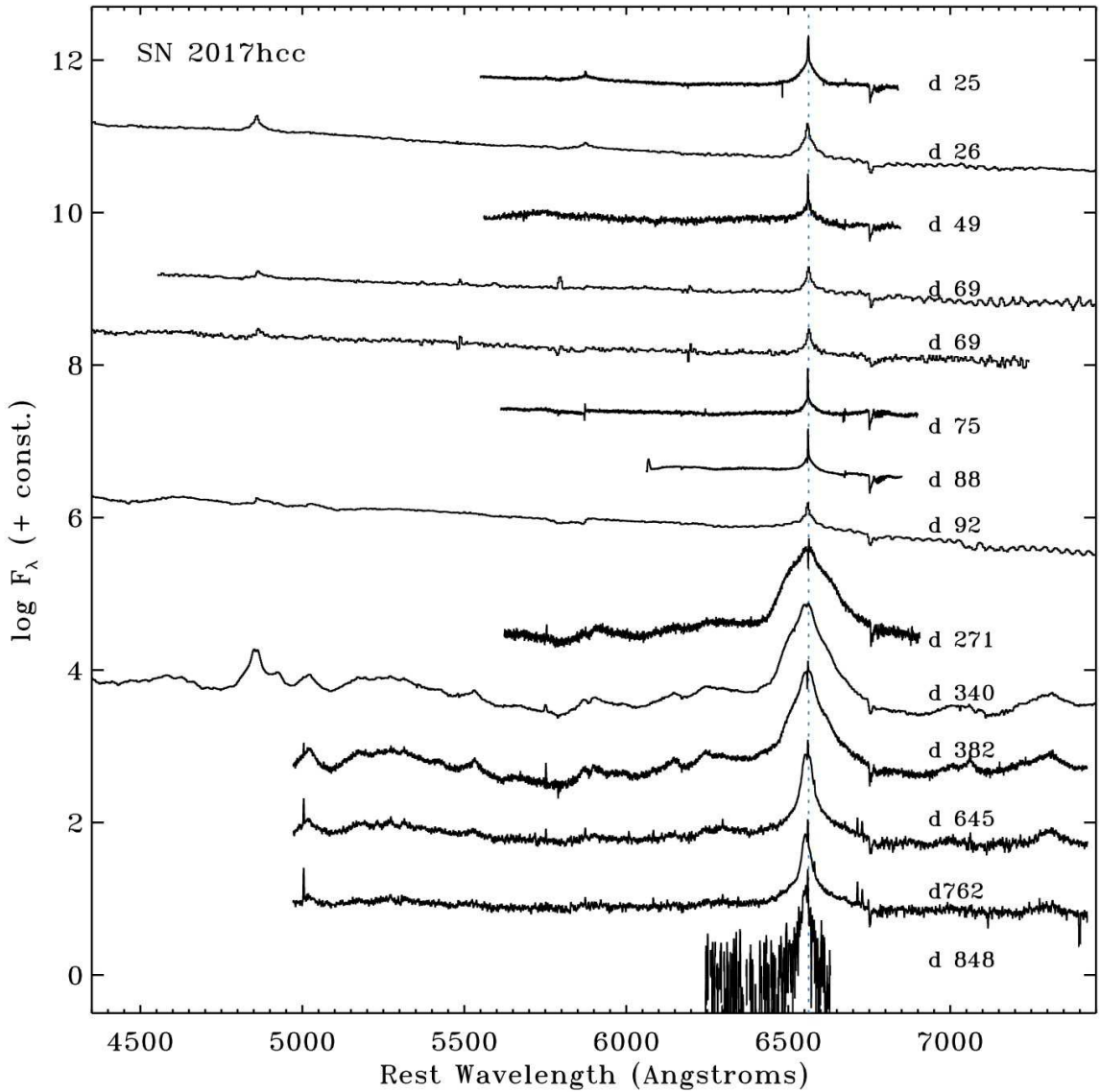}
\caption{Sequence of low-resolution optical spectra from the
  MMT, Bok, and Magellan (see Table~1). Flux is on a log scale.}
\label{fig:spec}
\end{figure*}

\subsection{Low-resolution optical spectra}

We obtained spectra of SN~2017hcc using the 6.5-m Multiple Mirror
Telescope (MMT) with three different instruments, including the
Bluechannel (BC) spectrograph, the Redchannel (RC) spectrograph, and
the newly commissioned BinoSpec spectrograph \citep{fabricant19}.  Each MMT Bluechannel or
Redchannel observation was taken with a 1.0 arcsec slit and either the
1200 lines mm$^{-1}$ grating covering a range of approximately
5800–7000 \AA, or the 300 lines mm$^{-1}$ grating covering a range of
about 3600-8000 \AA.  Standard reductions were carried out using
IRAF\footnote{IRAF, the Image Reduction and Analysis Facility, is
  distributed by the National Optical Astronomy Observatory, which is
  operated by the Association of Universities for Research in
  Astronomy (AURA) under cooperative agreement with the National
  Science Foundation (NSF).}  including bias subtraction,
flat-fielding, and optimal extraction of the spectra. Flux calibration
was achieved using spectrophotometric standards observed at an airmass
similar to that of each science frame, and the resulting spectra were
median combined into a single 1D spectrum for each epoch.

We obtained several late epochs of visual-wavelength spectra using
Binospec on the MMT.  These data were all taken using the 600 
lines mm$^{-1}$ grating centered on 6300 \AA\, (covering a range of
roughly 5100--7500 \AA) and with a 1.0 arcsec slit.  All data were
reduced using the Binospec pipeline \citep{Kansky2019}, which includes
an internal flux calibration into relative flux units from throughput
measurements of spectrophotometric standard stars.

We obtained a few epochs of low-resolution optical spectra with the
Boller \& Chivens (B\&C) spectrograph on the 2.3 m Bok telescope.  We also obtained one late-time spectrum using the
Inamori-Magellan Areal Camera and Spectrograph (IMACS;
\citealt{dressler11}) mounted on the 6.5 m Baade telescope of the
Magellan Observatory.  Data reduction for these followed standard
reduction for point sources in long-slit optical spectra, as
above.  Our low/moderate-resolution spectroscopic observations are
summarized in Table~\ref{tab:spectab1}, and all are plotted in
Figure~\ref{fig:spec}.

\subsection{MIKE echelle spectra}

We observed SN~2017hcc on three separate occasions using the Magellan
Inamori Kyocera Echelle (MIKE), which is a double echelle spectrograph
designed for use at the Magellan Telescopes at Las Campanas
Observatory in Chile \citep{bernstein03}.  We obtained observations of
SN~2017hcc during its main luminosity peak on 2017 Oct 25 UT (a total
exposure time of 2700 sec), and at two later epochs after it faded by
several magnitudes on 2018 Jul 11 and Sep 17 UT (total exposure times
of 5400 sec and 3600 sec, respectively).  All three observations had
good transparency and seeing (roughly 0.8 arcsec or better), and all
had the 1 arcsec $\times$ 5 arcsec slit aperture aligned at the
parallactic angle.  This yields a resolving power
$R$=$\lambda$/$\Delta\lambda$ of roughly 30,000, or a resolution of
typically 10 km s$^{-1}$ (although we note that the achieved
resolution measured from sky lines was somewhat narrower than this,
about 8 km s$^{-1}$, because the seeing was better than the slit width
on all three epochs).

\begin{figure}
\includegraphics[width=3.1in]{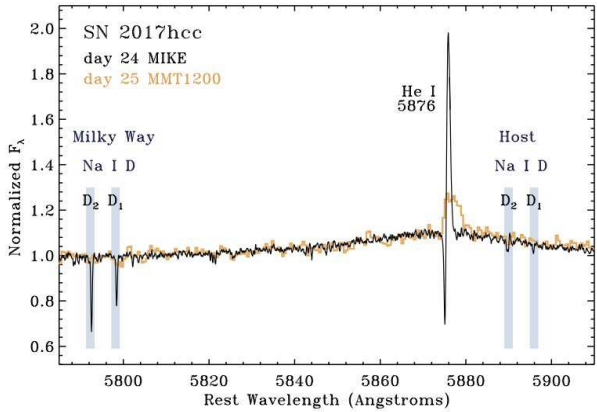}
\caption{MIKE spectrum on day 24 in the region around Na~{\sc i} D and
  He~{\sc i}~$\lambda$5876, shown in black.  The moderate-resolution
  MMT/Bluechannel spectrum taken one day later with the 1200 lines
  mm$^{-1}$ grating is shown in orange for comparison.  Expected
  locations of the D$_1$ and D$_2$ lines of the doublet are indicated
  by vertical blue bars for the redshift-corrected host galaxy and
  Milky Way components.  Narrow P Cygni absorption
  of He~{\sc i} $\lambda$5876 is strong in the echelle
  spectrum, but is washed out in the moderate resolution spectrum, and even
  the strong Milky Way Na~{\sc i} D absorption is barely detected in
  the MMT  spectrum.}
\label{fig:naid}
\end{figure}

\begin{figure}
\includegraphics[width=3.3in]{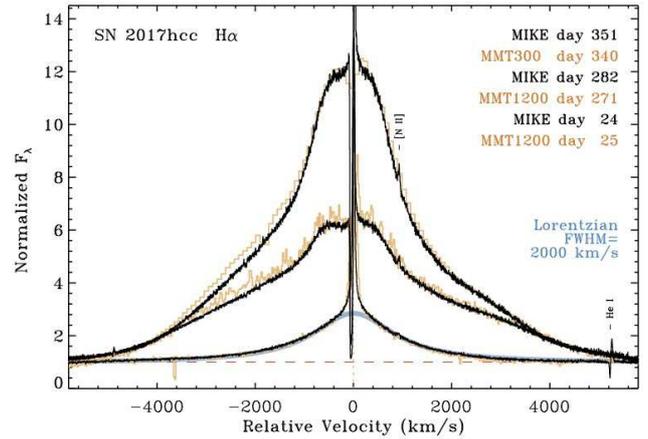}
\caption{Full line profiles of H$\alpha$ as seen in the three epochs
  of MIKE spectra (black).  The orange line profiles are from low- or
  moderate-resolution spectra taken around the same time, for
  comparison.  Because the H$\alpha$ line wings are so broad and the line
  center was located near the end of an echelle order, we used one
  order for blueshifted velocities and the adjacent order for the red
  wing of the line. The blue solid curve shows a symmetric Lorentzian
  profile for comparison with the early epoch line profile shape. The
  flux is on a linear scale.}
\label{fig:Haint}
\end{figure}

\begin{figure}
\includegraphics[width=3.3in]{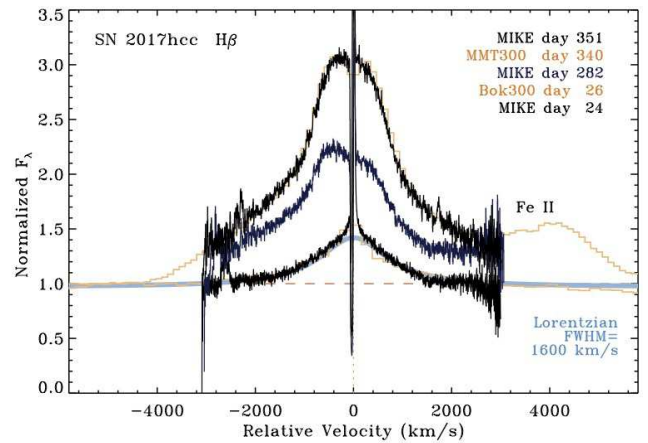}
\caption{Same as Figure~\ref{fig:Haint}, but for H$\beta$.  Zero
  velocity of H$\beta$ was located near the center of a MIKE echelle
  order, so we only show one full echelle order (there
  were slight flux mismatches and increased noise at the edges of
  adjacent orders. Aside from the edges
  of the orders, the line shapes in echelle spectra and lower
  resolution spectra are consistent. The bump at high positive
  velocity is emission from another line or lines (most likely Fe~{\sc
    ii}), rather than a strong excess of high velocity H$\beta$
  emission.  The flux is on a linear scale.}
\label{fig:HBint}
\end{figure}

\begin{figure}
\includegraphics[width=3.3in]{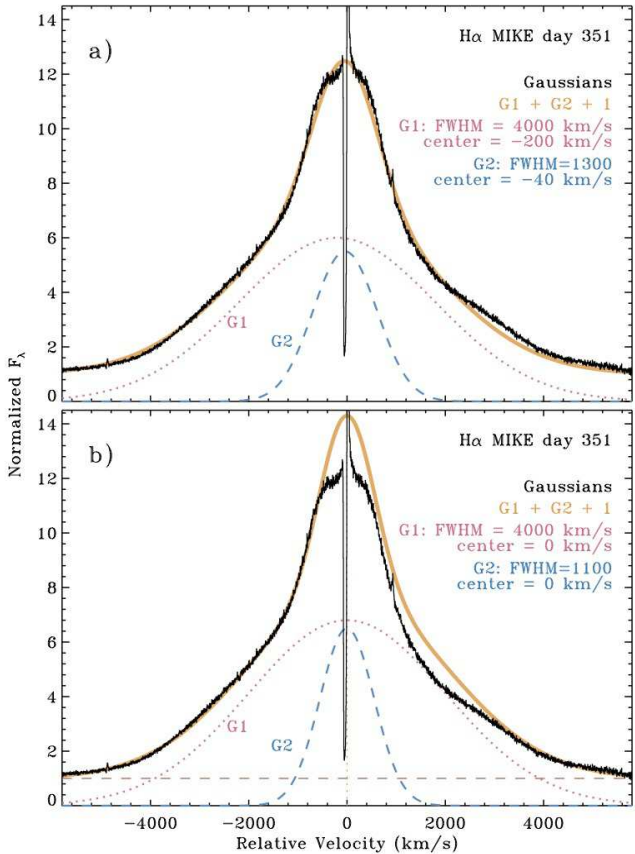}
\caption{Full line profile of H$\alpha$ as seen in the day 352 MIKE spectrum, compared to to different multicomponent
  Gaussians.  The top panel (a) shows two Gaussian profiles and their
  sum (plus continuum) where the centroids are allowed to be offset from zero, and the
  bottom (b) shows two similar Gaussians and their sum (plus continuum), but with the
  center fixed at 0 km s$^{-1}$, and where the sum is allowed to
  extend above the flux at low velocities and the red wing of the line
  (the excess flux may represent line flux lost from the intrinsic
  profile due to absorption).}
\label{fig:Hafits}
\end{figure}

The spectra were reduced using the latest version of the MIKE
pipeline\footnote{http://code.obs.carnegiescience.edu/mike/} 
(written by D. Kelson).
The reduced spectra were corrected for SN~2017hcc's redshift of $z$ =
0.01686, and each epoch had velocities converted to the heliocentric
reference frame (this correction matters for the narrow components
from the pre-shock CSM).  Figure~\ref{fig:naid} shows the region of
the spectrum including the Na~{\sc i} D doublet useful for evaluating
interstellar extinction and reddening, and also includes He~{\sc
  i}~$\lambda$5876 emission from the SN.  Figures~\ref{fig:Haint} and
\ref{fig:HBint} show the full line profiles of H$\alpha$ and H$\beta$,
respectively.  For H$\alpha$, the line center was located near the
edge of two adjacent echelle orders, so to display the full profile in
Figure~\ref{fig:Haint}, we spliced together two adjacent echelle
orders.  Because of the drop in sensitivity and the large throughput
corrections needed at the ends of echelle orders, we checked the
resulting overall shape of the line profile by comparing it to
lower-resolution single-order spectra taken very close in time (see
above).  These are plotted in orange in Figure~\ref{fig:Haint},
showing very good agreement in overall line shape.  Only a single
echelle order is plotted for H$\beta$, because it was centered in the middle
of an ehelle order.  For H$\beta$, we caution that the red wing of the
line is partly blended with Fe~{\sc ii} lines, so this should not be
interpreted as excess redshifted H$\beta$ flux.  For both H$\alpha$
and H$\beta$, the first epoch line profile is compared to a Lorentzian
line profile shape (thick light-blue curve), meant to match the line
wings with FWHM = 2000 km s$^{-1}$ for H$\alpha$ and 1600 km s$^{-1}$
for H$\beta$.  Figure~\ref{fig:Hafits} shows examples of Gaussian
components that may be fit to the day 351 H$\alpha$ line profile
shape, discussed in more detail later.

\begin{figure}
\includegraphics[width=2.9in]{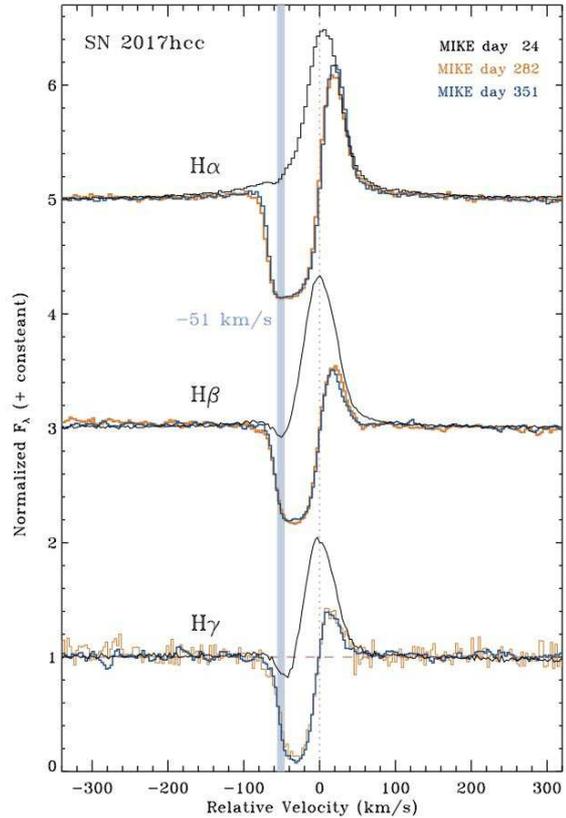}
\caption{Detail showing the narrow components of the brightest Balmer
  lines H$\alpha$, H$\beta$, and H$\gamma$ in the three epochs of
  echelle spectra.  For the first epoch of each, the observed profile was
  divided by an underlying broad Lorentzian (blue in
  Figure~\ref{fig:Haint}).  For the two later epochs, the ``continuum''
  for normalization was the intensity level of the plateau in the
  underlying broad/intermediate-width line profile. The intensity is on a
  linear scale.}
\label{fig:Hanar}
\end{figure}

\begin{figure}
\includegraphics[width=2.9in]{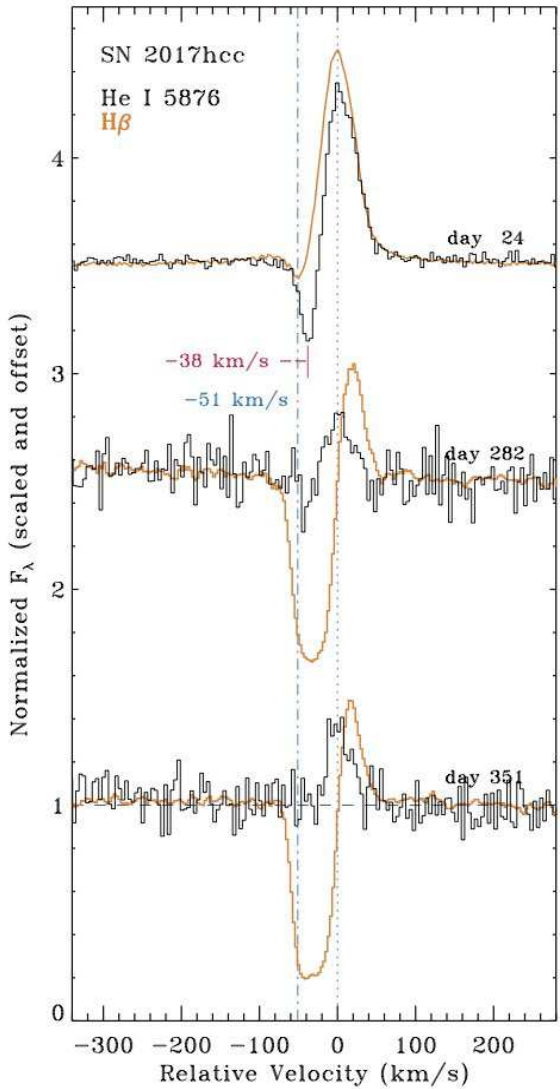}
\caption{Detail of MIKE echelle spectra on the three epochs showing
  the narrow component of He~{\sc i} $\lambda$5876 (black) compared to
  H$\beta$ (orange).  Approximate velocities of the early-time narrow
  absorption trough of H$\beta$ (at $-$51 km s$^{-1}$; blue) and
  He~{\sc i} (at $-$38 km s$^{-1}$; magenta) are marked.}
\label{fig:He5876nar}
\end{figure}

\begin{figure}
\includegraphics[width=2.9in]{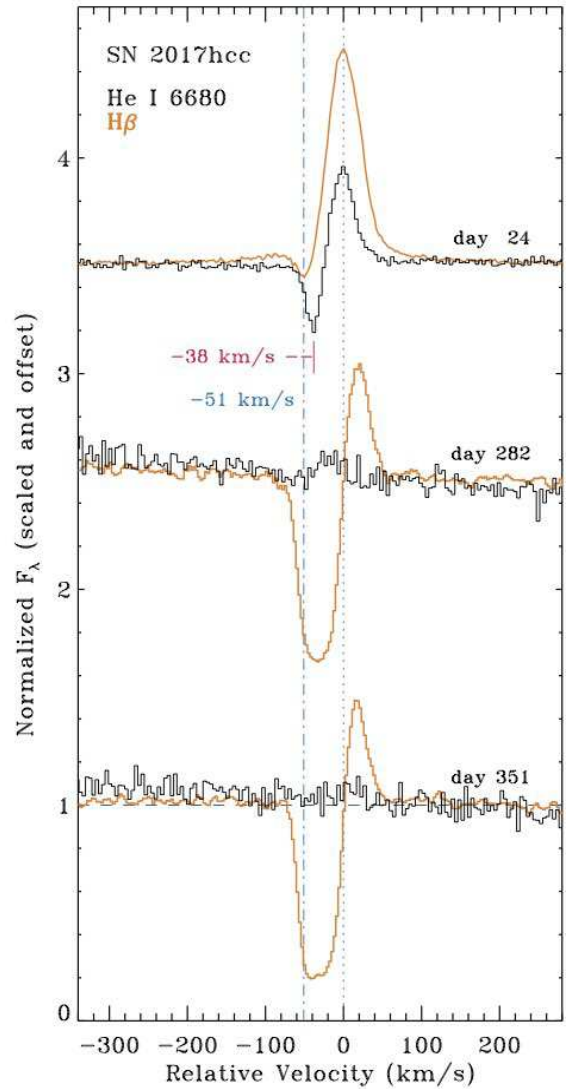}
\caption{Same as Figure~\ref{fig:He5876nar}, but for He~{\sc i} $\lambda$6680.}
\label{fig:He6680nar}
\end{figure}

Figure~\ref{fig:Hanar} zooms-in on the narrow components of H$\alpha$,
H$\beta$, and H$\gamma$ seen in MIKE spectra, showing the P~Cygni
profiles arising from the unshocked CSM.  For the first epoch, the
``continuum'' level is set at the continuum after subtraction of the
Lorentzian profiles shown in Figures~\ref{fig:Haint} and
\ref{fig:HBint}.  For the two later epochs, the ``continuum'' for
normalization refers to the flux level of the broader emission
component within $\pm$400 km s$^{-1}$. Similarly,
Figures~\ref{fig:He5876nar} and \ref{fig:He6680nar} zoom-in on the
narrow CSM components of He~{\sc i} $\lambda$5876 and $\lambda$6680,
respectively, with both being compared to the narrow components of
H$\beta$. (We do not show He~{\sc i} $\lambda$7065 because
SN~2017hcc's redshift caused this line to overlap with many narrow
telluric absorption lines, complicating the interpretation of narrow P
Cygni absorption features.)

\begin{figure}
\includegraphics[width=3.1in]{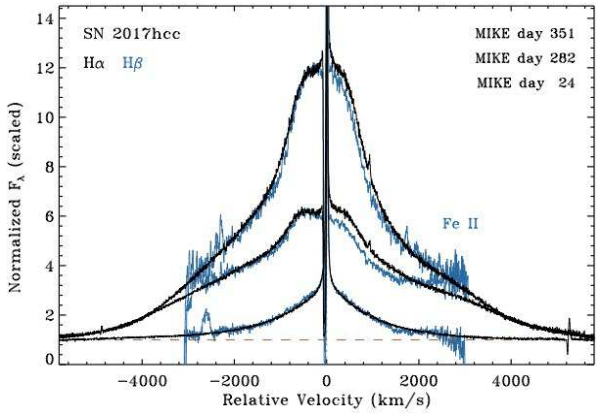}
\caption{Comparison between the full line profile shapes of H$\alpha$
  (black) and H$\beta$ (blue) in the three epochs of echelle spectra.
  The H$\beta$ strengths have been scaled to match H$\alpha$, to
  correct for their different line/continuum flux ratio, and so that
  their shape can be compared directly.  For the two later epochs,
  there appears to be a slight deficit of H$\beta$ flux on the red
  wing at intermediate velocities. The flux is on a linear scale.}
\label{fig:HaHb}
\end{figure}

Finally, Figure~\ref{fig:HaHb} compares the full line profile shapes
of H$\alpha$ (black) and H$\beta$ (blue) superposed on one another.
We have chosen to align these by scaling the line flux to match the
shape of the blue wing of the emission line, in order to determine if
there are any differences in the shape of the red wing.  The H$\beta$
line strengths have therefore been scaled arbitrarily in flux above
the continuum level.  For the first early epoch, the profile shapes
of H$\alpha$ and H$\beta$ agree remarkably well.  For the later two
epochs, however, the red wings of H$\beta$ are slightly depressed
compared to H$\alpha$.  This indicates that there is a wavelength
dependence to the line shape that we discuss later.

\subsection{Near-IR spectra}

\begin{figure}
\includegraphics[width=2.9in]{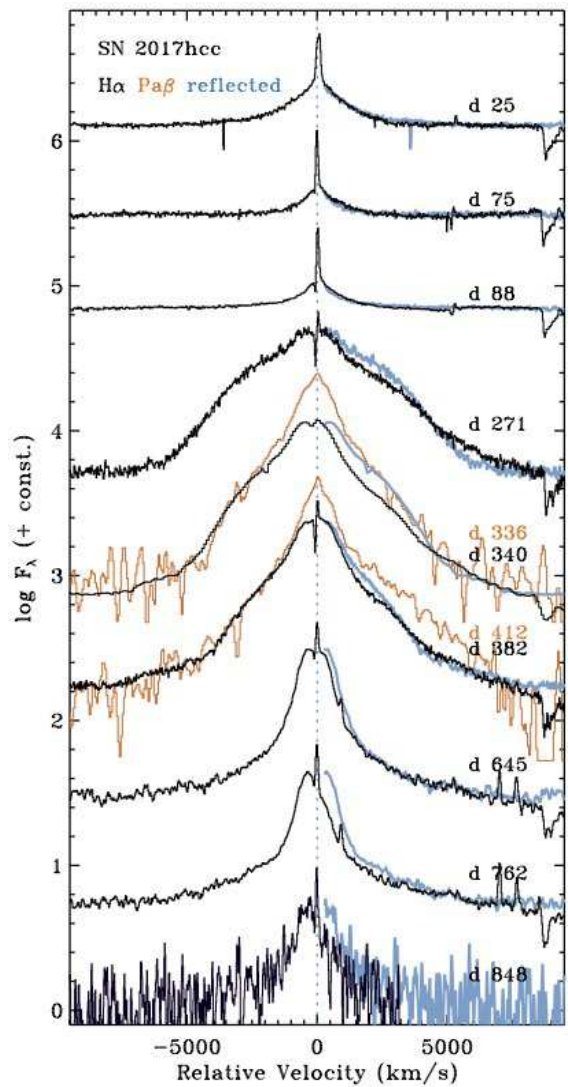}
\caption{Time series of the H$\alpha$ line profile shape in our
  moderate-resolution spectra (MMT Bluechannel and Binospec).  For
  each date, the black line is the observed line profile, whereas the
  blue line shows the blue wing of the emission line reflected across
  v=0 km s$^{-1}$.  This is meant to show how the red wing of the line
  would appear if the line were symmetric.  The orange spectra are the
  infrared Pa$\beta$ line from our MMT/MMIRS spectra on days 336 and
  412, for comparison with the H$\alpha$ profile at similar
  epochs.  The Pa$\beta$ profile is expected to suffer less
  extinction, and its red wing roughly matches the blue reflected
  wing of H$\alpha$.  Pa$\beta$ indicates that significant
  line flux may be absorbed around zero velocity in optical lines.
  The flux is on a log scale.}
\label{fig:Halow}
\end{figure}

We obtained two epochs of near-IR spectra covering the $J$, $H$, and
$K$, bands (roughly 1$-$2.3 $\mu$m) using the MMT and Magellan
Infrared Spectrograph \citep[MMIRS;][]{McLeod2012} mounted on the MMT, with
observations listed in Table~\ref{tab:spectab1}.  The
standard long-slit zJ and HK single-order spectra were reduced using
the MMIRS data reduction pipeline \citep{Chilingarian2013}. In this paper, we are most
interested in the Pa$\beta$ 1.28 $\mu$m line profile shape,
plotted along with H$\alpha$ line profiles in Figure~\ref{fig:Halow}.

%\begin{figure}
%\includegraphics[width=3.3in]{../SPECTRA/plotspecHa.nar.eps}
%\caption{Detail showing the narrow component of H$\alpha$ in the three
%  epochs of echelle spectra.  For the first epoch, the observed
%  profile was divided by underlying broad Lorentzian profile (blue in
%  Figure~\ref{fig:Haint}.  For the two later epochs, the ``continuum''
%  for normalization was the intensity level of the plateau in the
%  underlying broad/intermediate-width line profile. The flux is on a
%  linear scale.}
%\label{fig:Hanar}
%\end{figure}

%\begin{figure}
%\includegraphics[width=3.3in]{../SPECTRA/plotspecHb.nar.eps}
%\caption{Same as Figure~\ref{fig:Hanar}, but for H$\beta$. The flux is
%  on a linear scale.}
%\label{fig:HBnar}
%\end{figure}

\begin{figure}
\includegraphics[width=3.0in]{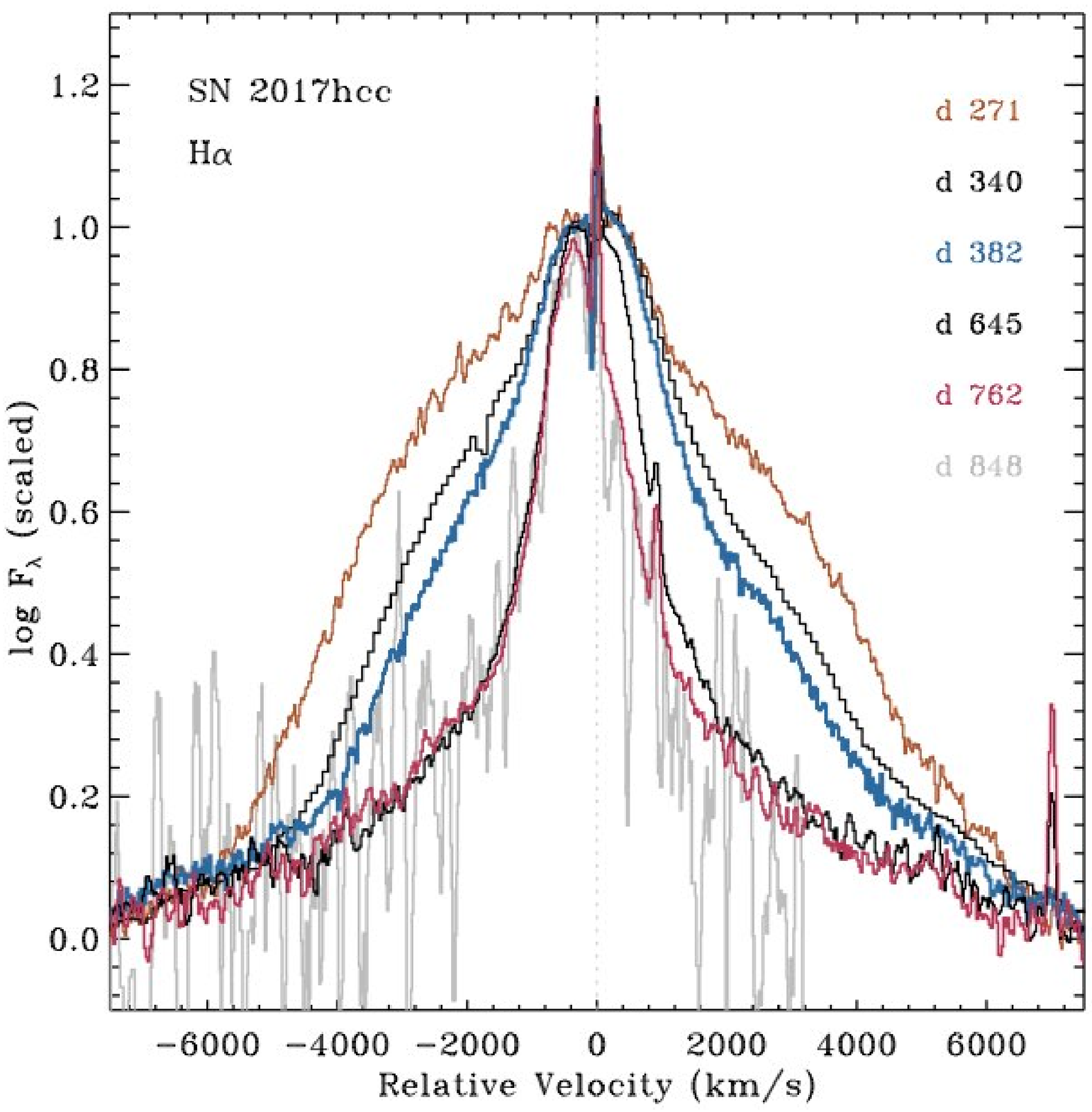}
\caption{Same as Figure~\ref{fig:Halow}, but with the late-time
  H$\alpha$ profiles plotted on top of one another, scaled arbitrarily to
  directly compare line shape.  Here we only show the late-time
  profiles after the symmetric Lorentzian scattering profiles no
  longer dominate the line shape, instead showing the broad component
  from the SN ejecta and the intermediate-width component from the
  CDS.  The flux is on a log scale.}
\label{fig:Haover}
\end{figure}

\section{RESULTS}

\subsection{Na I D and interstellar extinction}

The high-resolution MIKE echelle spectra allow a sensitive probe of
the interstellar absorption components from the Na~{\sc i} D doublet.
The individual D$_1$ and D$_2$ lines are easily resolved in the
echelle spectrum, and the high resolution allows greater
sensitivity to narrow absorption features, compared to
moderate-resolution spectra.  For examining interstellar absorption,
we consider early spectra when the continuum is strongest.
%\footnote{We do not detect the
%  5780 \AA\ diffuse interstellar band (DIB), but placing a reliable
%  upper limit on its equivalent width is complicated by possible broad
%  blueshifted He~{\sc i} $\lambda$5876.}

Figure~\ref{fig:naid} shows the region around Na~{\sc i} D, with
the day 24 echelle spectrum (black) compared to
the moderate-resolution spectrum taken
with MMT one day later (orange).  While both spectra have
enough resolution to separate the doublet, the absorption
features are more clearly defined in the MIKE spectrum.  One can
see the relatively strong Na~{\sc i} D absorption from the Milky Way
interstellar medium (ISM), which here is shifted to shorter
wavelengths because we are showing wavelengths corrected to the rest
frame of the host galaxy.  One can see the weaker Na~{\sc i}
D absorption from the ISM in the host galaxy.  In 
the lower resolution spectrum, the Milky
Way absorption is barely detected, and the host galaxy
absorption is undetected.  Figure~\ref{fig:naid} also shows the
strong narrow P~Cygni feature of He~{\sc i} $\lambda$5876 and its
wings, which will be discussed later.

For the Milky Way absorption component of Na~{\sc i} D, we measure 
equivalent widths of EW(D$_2$) = 0.129
($\pm$0.006) \AA, EW(D$_1$) = 0.085 ($\pm$0.005) \AA, and
EW(D$_2$+D$_1$) = 0.214 ($\pm$0.008) \AA.  Using the relations from
\citet{dovi12}, these translate to $E(B-V)$ values of 0.023 mag
(D$_2$), 0.028 mag (D$_1$), and 0.025 mag (D$_1$+D$_2$). As noted by
\citet{dovi12}, there is a 30-40\% systematic uncertainty in the
relation for $E(B-V)$.  These are in reasonable agreement
with the Milky Way line of sight reddening of $E(B-V)$=0.0285 mag
adopted earlier \citep{schlafly11}.

For the SN~2017hcc host galaxy ISM absorption component of Na~{\sc i}
D, we measure smaller equivalent widths of
EW(D$_2$) = 0.026 ($\pm$0.005) \AA, EW(D$_1$) = 0.014 ($\pm$0.004)
\AA, and EW(D$_2$+D$_1$) = 0.039 ($\pm$0.006) \AA.  Using the
relations from \citet{dovi12}, these would translate to $E(B-V)$
values of 0.014 mag (D$_2$), 0.019 mag (D$_1$), and 0.016 mag
(D$_1$+D$_2$).  A note of caution is warranted here, because
these host galaxy EW values we measure are below the range of EW
over which the relations from \citet{dovi12} were calibrated, which
extend down to 0.05 \AA, and there is some indication that
data deviate from the fit at the lowest EW values.\footnote{When extrapolated
to an EW of 0 \AA, the slope of the fit gives non-zero $E(B-V)$ values
of 0.12-0.17 mag, so $E(B-V)$ values for the smallest EWs are probably
overestimates in this relation.}  We therefore regard $E(B-V)$=0.016
mag as an upper limit to the host galaxy reddening for SN~2017hcc,
corresponding to $A_V$ $\la$ 0.05 mag.  This small host
reddening and extinction makes no difference in our analysis
that concentrates on line profile shapes, but this result may be useful for other
studies of the photometry and polarization of SN~2017hcc.

\subsection{Broad and intermediate-width components}

The line profiles of H$\alpha$ and H$\beta$ are characterized by
narrow emission with P Cygni absorption atop a
broader emission component.  At early times, the broader component has
a Lorentzian shape, with a FWHM value of about 2000 km s$^{-1}$ or
1600 km s$^{-1}$ for H$\alpha$ and H$\beta$, respectively (see
Figs.~\ref{fig:Haint} and \ref{fig:HBint}).  At later times, however,
the broader line shape becomes less symmetric.
The line shape at later epochs after day $\sim$100 is clearly
not a single Gaussian or Lorentzian shape, but seems to have at least
two subcomponents, with an intermediate-width component at velocities
below 2000 km s$^{-1}$, and a broader component extending out to
around 5,000 km s$^{-1}$ on the blue and red wings of the line.  The
line asymmetry permits multiple ways to
fit the line shape.

Figure \ref{fig:Hafits} shows two examples of how the same line might be
approximated with two Gaussian components.  Figure~\ref{fig:Hafits}a
(top) shows an example where the centroids of the Gaussians are
permitted to shift, and Figure~\ref{fig:Hafits}b (bottom) shows Gaussian components that have a center fixed at zero
velocity, but allowing red wings to fall below the model.

In Figure~\ref{fig:Hafits}a, even two broad components with FWHM
values of 4000 and 1300 km s$^{-1}$, with line centers shifted by
$-$200 and $-$40 km s$^{-1}$, respectively, are insufficient to give a
satisfying fit the detailed line shape.  The observed central
intermediate-width component is more boxy than the Gaussian, and the
broad wings do not match the Gaussian shape well.  An additional
broad Gaussian would be needed at roughly +3000 km s$^{-1}$ to
account for a red emission bump in excess of the fit in
Figure~\ref{fig:Hafits}a.

In Figure~\ref{fig:Hafits}b, the two broad components with FWHM values
of 4000 and 1100 km s$^{-1}$ and centers at 0 km s$^{-1}$ do not fit
the line profile either.  However, these particular Gaussians are
chosen because the blue wing and the far red wing are matched very
well.  The Gaussian model clearly exceeds the observed flux from
$-$500 to +3000 km s$^{-1}$, but this is by design ---  the motivation for allowing this
missing flux is that some of the intrinsic line profile may
be absorbed at some velocities.  The utility
of this seemingly poor fit and the ``missing'' flux will be
apparent later.

Overall, spectra at later times (after the Lorentzian profiles
transition to broader lines) seem to consistently require at least two
separate components in the broader line profile shape.  (1) A broad
component with FWHM  widths of 4000-6000 km s$^{-1}$.  This is broader and
stronger at first (days 100-300), and then narrower and fading
at later times (after day 300).  The evolution of the relative
strength and width of the broad component with time can be seen in
Figure~\ref{fig:Haover}.  (2) An intermediate-width component with a
FWHM of 1000-1500 km s$^{-1}$ appears after the Lorentzian
components fade, and persists until the latest epochs.  As
described later, we attribute the broad component to
emission from the unshocked SN ejecta, and the intermediate-width
component to shocked gas in the CDS.  This is the
typical interpretation of these features in many SNe IIn
\citep{smith17}.  Whether one prefers the Gaussians offset from zero
or the ones that fit the red side of the line poorly depend on the
interpretation of the line profile asymmetry.

In any case, the line profile asymmetry is rather mild at day 200-400,
but the asymmetry gets more severe at later times, like in the day 762
spectrum seen in Figure~\ref{fig:Halow}.  Interestingly, there appears
to be a significant change from day 645 to day 762.  The blue wing of
the line is virtually identical at these two epochs, but the red wing
changes drastically (examine Figures~\ref{fig:Halow} and
\ref{fig:Haover}), with the line becoming much more asymmetric and
suppressed on the red side.  The likely interpretation of this change
is discussed later.

\begin{figure}
\includegraphics[width=3.0in]{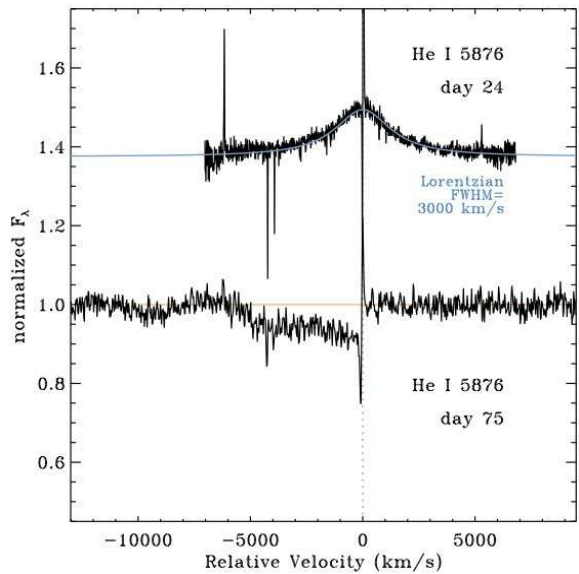}
\caption{The day 24 MIKE spectrum around He~{\sc i} $\lambda$5876, compared to a Lorentzian line profile (blue) with a FWHM = 3000 km s$^{-1}$, and the day 75 MMT spectrum showing the broad P~Cygni absorption seen in He~{\sc i} $\lambda$5876.  The weak absorption at around $-$10,000 km s$^{-1}$ on day 75 could be a fast component of He~{\sc i}, or may be from a different line.}
\label{fig:HeIbroad}
\end{figure}

The broader components of He~{\sc i} lines behave differently from Balmer lines.  At early times, He~{\sc i} $\lambda$5876 has a broad Lorentzian profile underneath the narrow P Cygni components, which can be seen in Figures~\ref{fig:naid} and \ref{fig:HeIbroad}.  This Lorentzian has a width of 3000 km s$^{-1}$, whick is faster/broader than the Balmer lines at the same epoch (this broader width may indicate electron scattering in hotter gas).  The broad He~{\sc i} emission is weak and fades during the first 100 days.  Interestingly, in our spectra, He~{\sc i} $\lambda$5876 shows the clearest evidence of broad blueshifted P Cygni absorption from fast ejecta during the main peak of the SN.  This broad absorption can be seen reaching to almost $-$6,000 km s$^{-1}$ in the day 75 spectrum (Figure~\ref{fig:HeIbroad}), and probably indicates that we are beginning to directly  see the fast SN ejecta at this epoch. (A weaker absorption feature can also be seen at $-$10.000 km s$^{-1}$, but this might arise from  different line transition.)  It is not so unusual to observe fairly strong broad blueshifted absorption from He~{\sc i} $\lambda$5876 in the ejecta of SNe~IIn during the main luminosity peak (e.g., \citealt{mauerhan13,smith+14}).  However, since Balmer lines still show strong Lorentzian profiles that indicate high electron scattering optical depths in the CSM at this same epoch, this suggests that we are able to see the fast SN ejecta because of asymmetric geometry.  For instance, we may be looking down on polar regions of the SN ejecta, despite high continuum optical depths in the equatorial CSM.

\subsection{Narrow CSM components}

Our high-resolution MIKE echelle spectra are particularly useful for
investigating the narrow emission and absorption arising from the
pre-shock CSM.  The narrow CSM lines are well resolved, appearing
significantly broader ($\sim$50 km s$^{-1}$) than the instrumental
resolution of $\la$10 km s$^{-1}$.

The narrow components of the three Balmer lines H$\alpha$, H$\beta$,
and H$\gamma$ are shown in Figure~\ref{fig:Hanar}.  All three lines
show qualitatively similar evolution, with a strong narrow emission and
weak P Cygni absorption at the first epoch on day 24, transitioning to
much deeper P Cygni absorption and weaker emission at the later two
epochs (days 282 and 351).

On day 24, the narrow P Cygni absorption from H$\alpha$ is weak and
poorly defined, but the narrow H$\beta$ and H$\gamma$ absorption is more
clear.  The centroid of the H$\beta$ absorption is found at $-$51 km
s$^{-1}$ ($\pm$1 km s$^{-1}$), indicated by the vertical light blue bar in
Figure~\ref{fig:Hanar}.  Interestingly, the speed of the H$\gamma$
absorption is a little slower, at roughly $-$47 km s$^{-1}$ ($\pm$1~km 
s$^{-1}$), while the H$\alpha$ absorption (admittedly weak and harder 
to measure) seems to be at a faster speed of $-$55 
km~s$^{-1}$ ($\pm$2 km s$^{-1}$), so there
seems to be a march to the red for absorption components going
from H$\alpha$ to H$\gamma$.  There is a shift in
the opposite sense for emission components, with the peak marching
slightly to the blue as we go from H$\alpha$ to H$\gamma$.  The FWHM
of the emission components also gets narrower as we proceed
up the Balmer series, with FWHM values of 50  ($\pm$1 km s$^{-1}$), 
49 ($\pm$1 km s$^{-1}$), and 46 ($\pm$1) km~s$^{-1}$ for H$\alpha$, 
H$\beta$, and H$\gamma$, respectively.  Thus,
all three trends (shifts in absorption minimum, peak emission,
and FWHM) {\it trace lower outflow velocities for higher order Balmer
lines}.  (We note, however, that the decrease in FWHM may be partly
caused by the increase in strength of the P Cygni absorption from
H$\alpha$ to H$\gamma$, where the stronger P Cygni features decrease
the flux on the blue side of the emission component.)

There is also a change in the centroid of the P Cygni absorption to
lower velocities at later times in Balmer lines
(Figure~\ref{fig:Hanar}), but it is not a simple shift of the
absorption to slower speeds.  Rather, on days 282 and 351, the blue
edge of the P Cyg absorption stays roughly the same as in the first
epoch, but the absorption gets {\it wider} for all three Balmer lines.
It therefore appears that the absorption at later times is tracing a
larger range in CSM expansion speeds along the line(s) of sight,
including additional slowly expanding gas that was not seen in the
first epoch, rather than a net shift to lower speeds in the CSM.  The
widening absorption at later epochs also eats into the emission peak,
pushing it further to the red and making it weaker for all three
lines.

The narrow CSM components of He~{\sc i} lines show some interesting
differences compared to Balmer lines.  Figure~\ref{fig:He5876nar}
shows narrow components of He~{\sc i} $\lambda$5876 compared to
H$\beta$, while Figure~\ref{fig:He6680nar} shows the same for He~{\sc
  i} $\lambda$6680.  The day 24 spectra reveal well-defined narrow
absorption from He~{\sc i} lines, but it is at a slower outflow speed
than Balmer lines.  Both He~{\sc i} lines have an absorption minimum
at $-$38 ($\pm$1) km s$^{-1}$, more than 10 km s$^{-1}$ slower than H$\beta$.
The He~{\sc i} lines therefore seem to extend the trend of marching to
slower speeds at higher excitation and ionization.  Unlike the shift
at late times in Balmer lines, this is not a widening of the
absorption, but a clear shift of the narrow component to slower
outflow speeds. Also unlike Balmer lines, He~{\sc i} lines do not
develop much deeper and broader absorption at late times, but instead,
the absorption gets weaker (there seems to be weak He~{\sc i}
$\lambda$5876 absorption at about the same velocity of $-$38 km
s$^{-1}$ on day 282) or the lines are not detected.  There is some
lingering emission in He~{\sc i} $\lambda$5876 at both later epochs,
and only very weak emission on day 282 for He~{\sc i} $\lambda$6680.
The physical significance and interpretation of the CSM emission will
be discussed below in Section 4.

\subsection{Asymmetry and wavelength dependence}  

Above in Section 3.2 we noted a mild asymmetry in the H$\alpha$
profile at later epochs, requiring either a slightly blueshifted
centroid for Gaussian fits, or a deficit of flux on the red wing
compared to symmetric profiles (Figure~\ref{fig:Hafits}).  There is
also a subtle wavelength dependence and a time dependence to this
asymmetry that we discuss in more detail here.

Figure~\ref{fig:HaHb} shows the same MIKE spectra of H$\alpha$ and
H$\beta$ that appeared earlier, but here they are plotted
together.  The day 24 profiles of H$\alpha$ and H$\beta$ are nearly
identical, and both are consistent with symmetric Lorentzian profiles.  
The later spectra on days 282 and 351 show clear
differences in the line profile shapes.  The
blue wings of the lines match quite well at the later two epochs;
while we have admittedly scaled the line strengths to overlap on the
blue side, it is also true that the {\it shape} of the blue wings
match for H$\alpha$ and H$\beta$.  In contrast, the red sides of the
H$\alpha$ and H$\beta$ lines do not match on days 282 and 351, with
H$\beta$ showing a clear deficit of emission on the red wing of the
intermediate-width component on both dates.  The
blueshifted asymmetry in the H$\alpha$ profile becomes more pronounced
at H$\beta$.

A consistent extension of this trend in the wavelength dependence is
seen as we move to the IR.  Figure~\ref{fig:Halow} includes a time
series of H$\alpha$ line profiles in low-resolution spectra, but it
also includes the 1.28 $\mu$m Pa$\beta$ line from our two late epochs
of IR spectra taken with MMIRS on the MMT.  Comparing H$\alpha$ and
Pa$\beta$, it is evident that these profiles have very different
shapes.  If we match the flux on the blue side of the line profile,
then Pa$\beta$ shows an excess in the peak of the
intermediate-width component (within $\pm$800 km s$^{-1}$) and
excess flux over most of the red wing of the line.  Interestingly, the
excess of Pa$\beta$ over H$\alpha$ in Figure~\ref{fig:Halow} is quite
similar to the excess of a symmetric Gaussian model above the observed
H$\alpha$ profile shown in Figure~\ref{fig:Hafits}b.  Overall, we
conclude that there is a subtle but clear wavelength dependence in the
H emission lines, such that lines at shorter wavelengths have a more
pronounced blueshift.

The asymmetry in emission profiles is also time dependent. 
Figure~\ref{fig:Halow} also shows the blue wing of the line reflected
across to the red side of the profile (in light blue) in order to
illustrate deviations from a symmetric profile shape.  We can
characterize the evolution in three main phases:

1. Early times (up to day 100) show little asymmetry, with a
narrow component atop broader symmetric Lorenztian profiles.  Small 
deviations from symmetry are that days 75
and 88 seem to show some excess flux on the red side (i.e. in the
opposite sense of the blueshifted asymmetry at later times).  This
might indicate some broad P~Cyg absorption that suppresses the blue wing of H$\alpha$ (recall that broad blueshifted He~{\sc i} is clearly seen at this epoch) or a
slight redward shift in the centroid of the Lorentzian profile.

2.  Intermediate phases (roughly days 200-400 sampled by our spectra)
show both a broad component and an intermediate-width
component.  These epochs exhibit a significant
deficit of flux on the red side of the line in both the
intermediate-width and broad components, although the deviation from
symmetry changes from one epoch to the next.  This asymmetry is
wavelength dependent, with more pronounced blueshift
at shorter wavelengths.

3.  Late phases (after day 500-600 or so) are dominatedf by an
intermediate-width component, with the broad component having faded
substantially or become narrower so that it is blended with the
intermediate-width component.  During this late phase, the
intermediate-width component shows a clear blueshifted asymmetry that
becomes more pronounced with time.  Note the discrepancy between the
observed red wing compared to a reflected blue wing, moving from day 
645 to 762 (Figure~\ref{fig:Halow}).

The profile in the day 762 spectrum
has a striking asymmetry, with a peak shifted
to about $-$350 km s$^{-1}$.  This profile cannot be fit by a
symmetric Gaussian or Lorentzian that has a shifted centroid.
Rather, the peak is skewed to the blue, missing emission at low
redshifted velocities, though the broader wings are almost
symmetric.

The change from phase 2 to 3 is even more evident in
Figure~\ref{fig:Haover}, where we overplot the H$\alpha$ profiles at
various times (excluding the early Lorentzian phase).  Here we see
that the broad component decreases in strength and/or width from day
271 to days 340 and 382, and then the broad component is gone by days
645, 762, and 848, while the width of the intermediate component
changes very little except for the increase in blueshifted asymmetry
of the peak.

\section{DISCUSSION}

In the discussion below, numbered days refer to days since the
inferred explosion date of 2017 Oct 1 (see above).  On this scale, for
reference, the time of peak visual light was around day 40-45, and
the peak bolometric luminosity was around day 30 \citep{prieto17}.

\subsection{Overall Line-profile Evolution}

SN~2017hcc exhibits the classic evolution of line
profile shapes that is common in strongly interacting SNe~IIn, which
transition from symmetric Lorentzian profiles at early times
(before and during peak), to irregular, broader,
and asymmetric shapes at late times well after peak.  This is
understood as a shift from narrow CSM lines broadened by electron
scattering to emission lines formed in the post-shock
CDS \citep{smith17,smith08tf,dessart15}.

The early profiles are characterized by a very narrow emission-line
core (about $\sim$50 km s$^{-1`}$) that also has a narrow P Cygni
absorption component.  These early profiles have broad wings that
follow a symmetric Lorentzian shape, due to incoherent electron
scattering of narrow emission from pre-shock gas
\citep{chugai01,chugai18,smith08tf,smith10,smith12,smith17}.
At these times, the narrow line width traces the pre-SN mass-loss
speed, whereas the line wings are caused by thermal broadening and are
not due to expansion speeds.  This indicates that at these early times
(usually up to and including the time of peak luminosity), the
continuum photosphere is in the CSM ahead of the shock, hiding
emission from the CDS and SN ejecta.  By day 65, however, we may begin to see the fast SN ejecta directly via broad P Cygni absorption in He~{\sc i} $\lambda$5876 (Figure~\ref{fig:HeIbroad}).  Seeing this broad absorption while still seeing Lorentzian profiles in Balmer lines may require asymmetry in the CSM.

At later times, however, emission from the fast SN ejecta and CDS become visible after
optical depths drop and the photosphere recedes (in mass), and as the
shock overtakes the photosphere.  For SN~2017hcc, this transition took
place sometime during a gap in our spectral coverage between days 92
and 271, missed because SN~2017hcc was behind the Sun.  From
day 271 onward, spectra reveal a more complex line profile shape in
H$\alpha$ with at least two broader components
(Figure~\ref{fig:Hafits}).  These two include a broad emission
component with FWHM = 4000 km s$^{-1}$ that we interpret as tracing
the fast, unshocked SN ejecta, as well as an intermediate-width
component with FWHM = 1100 km s$^{-1}$, which we interpret as emission
from the post-shock gas in the CDS.\footnote{One might infer that the ``broad'' width of 4,000 km s$^{-1}$ is not so fast when compared to typical speeds of $\ga$10,000 km s$^{-1}$ in the SN ejecta of non-interacting SNe.  However, recall that we are seeing these broad lines at late times after day 200, when broad lines are long gone in normal SNe~II-P, and only narrow nebular lines from the inner ejecta remain.  In this context, the longevity of lines with widths of 4,000 km s$^{-1}$ in SN~2017hcc is remarkable.} Narrow emission and P~Cyg
absorption from the pre-shock CSM persist to late times as well.
%Interpreting the SN ejecta speed and kinetic energy from the 4000 km
%s$^{-1}$ FWHM after day 271 is complicated.  There most likely were
%much faster SN ejecta that would have been seen at earlier times in a
%normal SN if their emission was not hidden by the optically thick CSM.
%On the other hand, these lines may also be broader than typically seen
%in non-interacting SNe at late times because radiation from the CSM
%interaction itself may illuminate SN ejecta that would normally have
%recombined and become transparent.

This transition is typically seen in SNe~IIn
\citep{smith08tf,smith17}, and essential properties of the
transition are reproduced in radiative transfer simulations of
SNe~IIn \citep{dessart15}. These simulations affirm the
interpretation of electron scattering in the CSM at early times and
emission from post-shock gas at later times.

An interesting aspect of the H$\alpha$ line-profile evolution in
SN~2017hcc is the clear identification of a broad emission component
reaching $\pm$6,000 km s$^{-1}$ that we attribute to the fast SN
ejecta (Figures~\ref{fig:Hafits}b, \ref{fig:Halow}, and
\ref{fig:Haover}).  After day $\sim$200, this broad component has
comparable strength to the intermediate-width component from the
post-shock CDS and it dominates the appearance of the H$\alpha$ line.
As time progresses, the broad component declines in strength
and width and disappears by day 752, whereas the intermediate-width
component persists at all late epochs with a roughly constant width
(ignoring affects associated with asymmetric absorption; see below).
This different time evolution confirms that the two emission
components have a different origin from one another. Seeing strong
emission from the freely expanding SN ejecta is rare in SLSNe IIn,
where continuum optical depths often hide the emission from underlying
ejecta, or where stronger emission from the intermediate-width
component dominates the lines.  For example, the
broader component from SN ejecta was absent or much weaker in
day$>$200 spectra of SN~2010jl \citep{smith12,gall14,fransson14}, and
SN~2006tf showed only weak H$\alpha$ emission and faint O~{\sc i} and
He~{\sc i} absorption at fast blueshifts
\citep{smith08tf}.  Since we see the SN ejecta more clearly in
SN~2017hcc, we may be viewing from a different
orientation (looking from the poles, for example).

Identifying the broad emission with the SN ejecta also has important
implications for interpreting the asymmetry in SN~2017hcc's line
profiles discussed below, and for its high observed continuum
polarization.  The broad emission component of H$\alpha$ shows only
mild asymmetry (Figure~\ref{fig:Haover}), with a small portion of its
red wing depressed compared to a symmetric profile
(Figure~\ref{fig:Hafits}).  Importantly, though, the red and blue
wings of the broad component are symmetric at velocities faster
than $\pm$3000 km s$^{-1}$.  Moreover, the line profile of the
infrared line Pa$\beta$, where any dust absorption should be less
influential than in the optical, is quite symmetric as well
(Figure~\ref{fig:Halow}).  This means that the intrinsic emission-line
profile from the fast SN ejecta is symmetric, which has two critical
implications: (1) High continuum optical depths associated with the SN
photosphere are not blocking the receding side of the explosion at
late times, because the broad lines are symmetric at the highest
velocities.  Therefore, something else is causing the blueshifted line
asymmetry in late-time spectra.  (2) The symmetric emission suggests
that the underlying SN explosion itself was not highly aspherical.
This, in turn, would mean that faster or denser SN ejecta in a
particular direction (i.e. a lopsided explosion) are not causing
stronger CSM interaction in a particular direction in SN~2017hcc, and asymmetry in
the underlying SN explosion is probably not responsible for
SN~2017hcc's high polarization.  Instead, the polarization is likely
related to aspherical CSM, and the blueshifted line profiles are
likely caused by selective absorption, discussed later.

\begin{figure}
\includegraphics[width=3.1in]{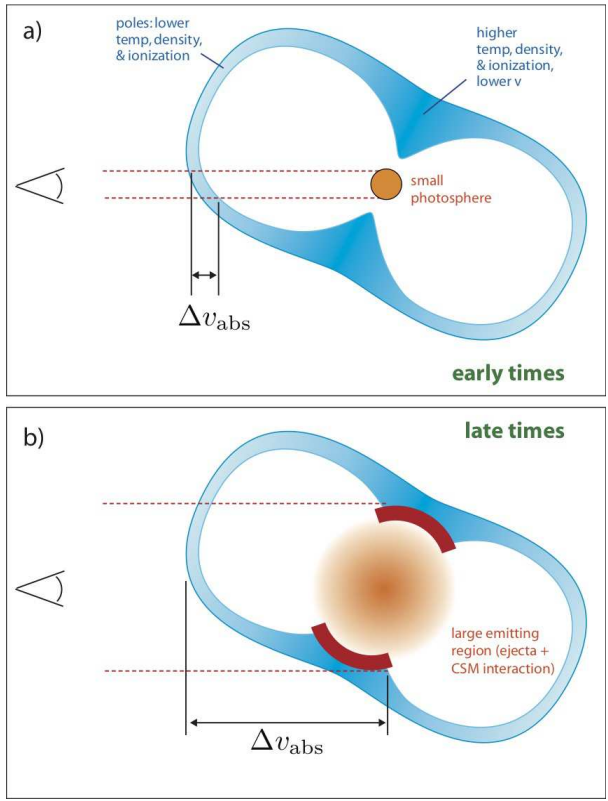}
\caption{Sketch illustrating how a hypothetical CSM geometry may
  explain the observed behavior of narrow emission and absorption
  lines in SN~2017hcc.  Panel (a) represents early times within the
  first $\sim$100 days, and panel (b) shows later times after day 200.
  The same bipolar CSM configuration (blue) is shown in both panels,
  where the structure is meant to symbolize latitude-dependent outflow
  velocity, rather than physical size and radius (although these are
  related if the CSM was made by an episodic mass ejection).  The main
  difference in the two panels is the change in the emitting source of
  continuum and broad/intermediate-width line emission: (a) at early
  times, the emitting source (orange) is the SN photosphere, which
  still has a relatively small radius.  This light passes through a
  small section of the polar CSM, resulting in a narrow range of
  absorbed velocities ($\Delta v_{\rm abs}$).  (b) at later times, the
  ejecta and shock have expanded, and most of the luminosity is
  generated in a more extended CSM interaction shock (dark red) and in
  the expanded SN ejecta (orange gradient).  This light passes through
  a much larger section of the CSM, yielding a broader range of
  $\Delta v_{\rm abs}$.  A hypothetical Earth-based observer is at
  left.  This observer sees narrow emission from all parts of the CSM
  shell, but only sees absorption generated along the line of sight to
  the continuum or broad-line emitting regions.}
\label{fig:sketch}
\end{figure}

\subsection{Narrow components and the pre-shock CSM}

\subsubsection{The narrow component at high resolution}

A novel aspect of this study is that we obtained three epochs of
high-resolution echelle spectra, which
provides a unique view of the narrow-line emission from slowly
expanding CSM.  Only a few examples of high-resolution echelle spectra
for SNe~IIn have been published, including SN~1998S on day 1
\citep{shivvers15}, SN~1997ab a few months after explosion
\citep{salamanca98}, SN~1997eg on roughly day 200 \citep{salamanca02},
and SN~2005gj (a Type~Ia/IIn hybrid) on days 86 and 374
\citep{trundle08}.  These had inferred progenitor wind speeds deduced
from narrow lines of 40 km s$^{-1}$ (SN~1998S), 90 km s$^{-1}$
(SN~1997ab), 160 km s$^{-1}$ (SN~1997eg) and 60-130 km s$^{-1}$
(SN~2005gj).

It is clear that with CSM expansion velocities of only 40-50 km
s$^{-1}$ in the case of SN~2017hcc, the CSM emission and absorption
are completely washed out in most low-resolution optical spectra that
are usually used for observing SNe (typically
$R$=$\lambda$/$\Delta\lambda$ of 300-1000, or up to 300 km s$^{-1}$
resolution for full broad wavelength range optical spectra).  The
narrow line profiles are underresolved even in moderate-resolution
spectra like those we typically obtain using a 1200 lpm grating with
Bluechannel on the MMT (Andrews et al., in prep.), or spectra obtained
with X-shooter on the VLT (typically $R$ of 4000-1000 or 30-80 km/s).
Examples of the narrow components in echelle spectra as compared to 
moderate resolution MMT ($R\sim$4000) spectra are shown in 
Figures~\ref{fig:naid}, \ref{fig:Haint}, and \ref{fig:HBint}.  
In the first epoch especially, the narrow P~Cyg absorption is lost in
moderate-resolution spectra.

Based on the results for narrow lines discussed above, we summarize
key observed properties for SN~2017hcc that any model for its CSM should 
explain:

1.  Early times show relatively weak blueshifted absorption with a
narrow range of speeds centered around 40-50 km s$^{-1}$.  The blue
edge of the absorption is around $-$60 or $-$70 km s$^{-1}$ (different
values for different lines).

2.  At later times, the blue edge velocity remains about the same, but
the range of absorbed speeds is wider, extending to slower
speeds, and the absorption is deeper.

3.  By day 75, we begin to see the fast SN ejecta, even though the optical 
depths in the CSM remain high.  This requires non-spherical geometry.

4.  Deeper Balmer line absorption at late times reaches down to
$\sim$20\% of the continuum, but this ``continuum'' is mostly
the underlying broad components of the same
emission line.  This means that the absorption source is
along our line of sight to both the photosphere and the
broad components.  The continuum luminosity
and intermediate-width component may arise in the
same CSM interaction region.

5.  The narrow emission (especially its red wing) stays roughly
the same at all epochs, and has a velocity width\ similar to
the early absorption speed around 40-50 km s$^{-1}$.  This
indicates that the asymmetry in CSM velocity is fairly mild, and that
the CSM speed ahead of the shock remains similar over a range of radii
as the shock moves out through it, regardless of the changing
luminosity of the SN.  This, in turn, suggests that any
non-sphericity in the CSM geometry is probably axisymmetric rather
than one-sided.

6.  The higher excitation and higher ionization lines, which are
normally formed deeper in a stellar wind or closer to the shock in a
SN, have slightly {\it lower} speeds (i.e. 38 km s$^{-1}$ in He~{\sc
  i} vs. 55 km s$^{-1}$ in H$\alpha$, and slightly decreasing speeds
in higher-order Balmer lines).  This same trend was also observed in 
narrow components of
SN~2005gj \citep{trundle08}.  We are not aware of any other SN~IIn
where this has been seen in high-resolution data. Interestingly, the
opposite was seen in spectra of SN~1998S, where the narrow cores of
higher-ionization lines were systematically {\it broader} than
lower-ionization lines \citep{shivvers15}.  This difference could be
partly due to the fact that the high-resolution spectrum of SN~1998S
was taken within 1 day of explosion, which still included emission
from the inner acceleration zone of the progenitor wind.

%%%%%%%%%%%%%%%%%%%%

One conjecture that is immediately apparent from this list is that a
single-speed, constant velocity spherical wind cannot reproduce these
properties.  Instead, some more complicated geometry or time
dependence is needed.

\subsubsection{Axisymmetric geometry in the CSM}

Figure~\ref{fig:sketch} shows a sketch that illustrates how a
hypothetical CSM geometry might account for the observed behavior of
narrow lines seen in SN~2017hcc.  While this is admittedly not a
unique explanation, it does account for several traits that a
spherical wind cannot.  The basic idea represented in this figure is
that the SN explodes inside a CSM shell that has a bipolar
configuration, borrowing from nebulae often seen around LBVs like
$\eta$ Car \citep{smith06,smith18alma}, except with much slower
expansion speeds than $\eta$~Car.  This slow CSM nebula (tinted blue
in Figure~\ref{fig:sketch}), viewed by an Earth-based observer at some
intermediate latitude, has higher velocities, larger radii, and a
larger inner cavity along the poles, and it has higher densities and
lower velocities in the equator.  As in the case of $\eta$~Car, such a
structure might arise from an LBV-like eruption in a massive binary
interaction or merger event \citep{smith18}.  The emission component
of the narrow lines arises in the portions of the CSM that are not
along the line-of-sight (again, the blue region in
Figure~\ref{fig:sketch}, and this emission retains approximately the
same velocity at all times because it is the integrated emission from
most of the CSM.

The key point conveyed in Figure~\ref{fig:sketch} is that the
underlying source of luminosity (tinted orange/red) is expanding with
time, causing the emitted radiation to traverse different paths (and
hence different velocity ranges) through the asymmetric CSM.  At early
times during the SN peak (Figure~\ref{fig:sketch}a), the emitting
radius is small compared to the size of the nebula, and so the
continuum radiation traces a pencil beam through only a small section
of the CSM.  In this case, it passes through only a small section of
the thin polar cap, yielding weak blueshifted absorption near the
maximum speed with only a narrow range of velocity (small $\Delta
v_{\rm abs}$).  At late times (Figure~\ref{fig:sketch}b; past day 200
when the continuum luminosity has dropped), the emitting source that
is absorbed becomes more complicated.  The continuum luminosity has
mostly faded, and the photons being absorbed by the CSM are now the
broad component from the freely expanding SN ejecta (orange gradient)
as well as the intermediate-width component from the strong CSM
interaction, which may be dominated by emission from the post-shock
CDS in the dense equatorial CSM (red arcs).  This broad line emission
passes through a much larger sample of the CSM and traces a wider
range of speeds (hence, larger observed $\Delta v_{\rm abs}$).  In
this case, the blue absorption edge is still about the same,
corresponding to the polar speed -- but absorption now occurs at
speeds all the way to zero, because some of the absorbing CSM is
moving transversely in the plane of the sky and has no Doppler shift.
The absorption is also deeper at these later times, because most of
the source photons pass through denser material and longer path
lengths through the low-latitude portions of the CSM.  Thus, this CSM
configuration meets the first 4 requirements listed above.

At some epoch intermediate between Figures~\ref{fig:sketch}a and \ref{fig:sketch}b, a transition occurs.  The photosphere is outside the shock and obscures the CDS and SN ejecta at early times, but eventually the CSM interaction at the equator becomes dominant, and the optical depths through polar regions thin first.  At this point, we may begin to see SN ejecta directly down the poles.  This may explain the broad blueshifted He~{\sc i} $\lambda$5876 absorption on day 75 (Figure~\ref{fig:HeIbroad}), while we can still see Lorentzian profiles arising from electron scattering of H emission lines in equatorial CSM regions.

The sixth requirement above, that slower velocities are seen in higher
ionization lines like He~{\sc i}, or higher order Balmer lines like
H$\gamma$, may be achieved in this configuration as follows.  These
lines will prefer regions of higher temperature and ionization.  In
the configuration shown in Figure~\ref{fig:sketch}, these are most
likely to be found closer to the source of SN luminosity and closer to
the strongest shock interaction - i.e., in the regions of the pinched
waist near the equator where the expansion velocities are slower and
directed out of our line of sight.  Lower velocities at smaller radii arise naturally
in CSM created in episodic mass loss, as opposed to a steady wind.
Thus, a bipolar configuration may naturally satisfy the requirement of
lower velocities for higher excitation.  This is harder to imagine in
a spherical CSM configuration.  At late times, the He~{\sc i} lines
may weaken significantly, as observed (Figures~\ref{fig:He5876nar} and
\ref{fig:He6680nar}), for a few reasons: this may happen because the
inner equatorial CSM has been swept up, or simply because the SN's
luminosity has dropped and the shock speed has slowed, allowing the
remaining pre-shock gas to cool and He to be mostly neutral.

%%%%%%%%%%%%%%%%%%

\subsubsection{Radiative acceleration or geometry?}

The sixth requirement above (slower speeds for higher excitation) also
has important consequences regarding any acceleration of pre-shock CSM
by the radiation from the SN itself \citep{chugai19}.  Radiative
acceleration of pre-shock CSM has sometimes been invoked to account
for blueshifted line profiles seen in some SNe~IIn, like SN~2010jl
\citep{fransson14,zhang12}.  However, this effect should be most important for
the peak-luminosity phase of the SN, whereas the blueshifted line
profiles persist to late times in most SNe~IIn.
Moreover, radiative transfer simulations for SN~2010jl
\citep{dessart15} indicate that radiative acceleration cannot
reproduce the observed blueshift.

For SN~2017hcc, the observed velocity patterns suggest that any effect
of pre-shock acceleration of the CSM is unimportant.  Radiation from
the shock could, in principle, propagate upstream so that photon
momentum from the tremendous SN luminosity could radiatively
accelerate the unshocked CSM, just as stellar luminosity radiatively
accelerates winds in massive stars.  If this happens in CSM with high
optical depth, as required in SNe~IIn, then we expect the strongest
pre-shock acceleration in upstream regions that are close to the
shock, tracing regions of higher temperature and ionization, and
milder or minimal acceleration at larger and cooler radii in the CSM.
Radiative transfer simulations confirm that H$\alpha$ arises at larger
radii than lines like H$\gamma$ or He~{\sc i}, which come from a
similar deeper region (e.g., \citealt{shivvers15,dessart15}). So in the case of
radiative acceleration by a SN~IIn, one expects higher
outflow speeds in He~{\sc i} lines or in H$\gamma$ as compared to
H$\alpha$, because these lines are closer to the shock.  This
effect was indeed seen in SN~1998S \citep{shivvers15}.  SN~2017hcc
shows the opposite trend, however, with {\it lower} outflow speeds in
He~{\sc i} and H$\gamma$.  This means that the observed differences in
speed are not tracing acceleration of the unshocked CSM by radiation
from the SN, and that any radiative acceleration of pre-shock CSM is
smaller than the 10 km s$^{-1}$ difference in these lines.  Radiative
acceleration of the pre-shock CSM must also, therefore, play no role
in the blueshifted asymmetry seen in emission lines at late times.
Instead, the likely explanation for the slower outflow speeds in
He~{\sc i} and H$\gamma$ may be geometric, as noted earlier.

%%%%%%%%%%%

The bipolar CSM geometry invoked in Figure~\ref{fig:sketch} is hardly
unusual; in fact, axisymmetric CSM tends to be the norm rather than
the exception among massive star nebulae
\citep{nota95,smith14,smith17}, probably due to the pervasiveness of
binary interaction in massive star evolution \citep{sana12,moe17}.
Overall, the CSM expansion around SN~2017hcc is fairly slow, with the
bulk speed of around 40-50 km s$^{-1}$ and a blue edge to the P~Cygni
absorption at only $-$70 km s$^{-1}$.  Because the absorption speed
seen at early times is comparable to the emission width, and because
the blue edge stays the same throughout its evolution, it is likely
that we view SN~2017hcc along a sightline corresponding to a
mid-latiutude or high latitude (say within 45$^{\circ}$ or so of the
symmetry axis).  This is different from SN~2010jl, where the P~Cyg
absorption speed is slower than the narrow emission width
\citep{smith11}, suggesting a view from low latitudes.

Although bipolar nebulae are common around LBVs \citep{nota95}, most
notably around $\eta$~Carinae \citep{smith06}, the expansion speed of
the CSM around SN~2017hcc is relatively slow compared to typical LBV
winds and nebular expansion speeds of $\ga$100 km s$^{-1}$
\citep{smith14}.  However, because of their bipolar geometry, some
LBVs exhibit a wide range of outflow speeds in a single object; for
example, although $\eta$~Car has a wind speed of 500 km s$^{-1}$
\citep{smith03,hillier06} and the poles of its nebula are expanding at
650 km s$^{-1}$ \citep{smith06}, it has much slower speeds of only 40
km s$^{-1}$ in its equatorial regions
\citep{zethson99,smith06,smith18alma}.  There are also several
well-studied blue supergiants that have very slow (10-40 km s$^{-1}$;
much slower than their stellar winds) expansion in their resolved
equatorial ring nebulae.  These include Sher~25 \citep{brandner97},
SBW1 \citep{smith13}, NaSt1 \citep{mauerhan15}, the massive eclipsing
binary RY~Scuti \citep{smith02,smith11ry}, HD~168625
\citep{smith07hd}, and of course the progenitor of SN~1987A
\citep{meaburn95,ch00}.  These slow disks are thought to arise from
binary interaction episodes, allowing outflows much slower than their
raditively driven winds.  The class of B[e] supergiants are also blue
supergiants that are inferred to have slow, dense, equatorial outflows
\citep{zickgraf06,kraus19}.  While the CSM around SN~2017hcc is
expanding faster than the winds of normal RSGs (typically 10-20 km
s$^{-1}$), there is a subset of extreme, high-luminosity RSGs with
faster winds and high mass-loss rates, some of which also have
asymmetric or axisymmetric structures in their CSM.  One pertinent
example is VY~CMa, which has mildly
bipolar geometry, with the bulk outflow at 35-40 km/s, but with some
faster features up to 70 km/s \citep{smith04,smith09,decin16}.
Another extreme RSG with bipolar geometry seen in water
masers is VX~Sgr \citep{berulis99,pashchenko06}.

%%%%%%

Similar bipolar/disk geometries, viewed from different
orintation angles, have been invoked for several other SNe with strong 
interaction, including SN~2009ip
\citep{mauerhan14,smith14}, SN~2010jl
\citep{andrews11,smith11,gall14,fransson14,dessart15}, SN~2007rt
\citep{trundle09}, PTF11iqb \citep{smith15}, SN~2014ab
\citep{bilinski20}, SN~2010jp \citep{smith12jp}, iPTF14hls
\citep{andrews18}, SN~2013L \citep{andrews17}, SN~2013ej
\citep{mauerhan17ej}, SN~1998S \citep{leonard00}, and SN~1997eg
\citep{hoffman08}, among others.  Common indications of bipolar
geometry and slow disks around SNe~IIn might hint at a common
mechansim related to pre-SN binary interaction \citep{sa14}.

\subsection{Pre-SN Mass Loss}

Armed with reliable estimates of the speeds of the CSM and CDS, and
the observed SN luminosity, we can make several rough estimates of the
pre-SN mass loss.  The observed narrow CSM lines indicate an outflow
speed of 40-50 km s$^{-1}$, so we take an average velocity for the
pre-SN mass loss of $V_{\rm CSM}$=45 km~s$^{-1}$.  The measured FWHM
of the intermediate-width component of H$\alpha$, emitted by the CDS,
is 1100 km s$^{-1}$.  We multiply this radial velocity by $\sqrt{2}$
to account for the expansion direction predominantly away from our
line of sight (see Figure~\ref{fig:sketch}), taking $V_{\rm CDS}$=1600
km s$^{-1}$.

The luminosity generated by CSM interaction (see \citealt{smith17} for
a review) is given by

\begin{equation}
L = \frac{1}{2} w V_{\rm CDS}^3
\end{equation}

\noindent where $w$ = $4 \pi R^2 \rho$ = $\dot{M}/V_{\rm CSM}$ is the
wind density parameter.  Here, $L$ is the total bolometric luminosity
generated by CSM interaction, which includes line emission and
continuum across all wavelengths.  The observed UV/optical/IR
continuum luminosity is a lower limit for this, but during the main
peak of the SN when the CSM interaction shock is below the photosphere
that resides in the optically thick CSM, the optical luminosity is a
good proxy for $L$.  At later times when the material becomes more
optically thin and the visual-wavelength luminosity drops, an
increasing fraction of the luminosity may escape as line emission and X-rays.
(The value of $L$ may also change as the SN evolves and
the shock decelerates or runs into varying density CSM.)  The
progenitor's mass-loss rate can be expressed as

\begin{equation}
\dot{M}_{\rm CSM} = 2 L \frac{V_{\rm CSM}}{V_{\rm CDS}^3}
\end{equation}

\noindent where this is likely a conservative value if $L$ is derived
from the visual-wavelength continuum luminosity, which is a lower limit to
the true bolometric $L$.  Again, the pre-SN mass-loss rate was eruptive
and episodic, so $\dot{M}_{\rm CSM}$ may not have been sustained for
very long, and only represents the density of the CSM into which the
SN shock expands during the main light curve peak.

As noted in the introduction, SN~2017hcc had a peak $V$ absolute
magnitude of about $-$20.7 mag, which translates to about
1.6$\times$10$^{10}$ $L_{\odot}$ or roughly 6.5$\times$10$^{43}$ erg
s$^{-1}$ with no bolometric correction.  With such a high luminosity,
it is unlikey that the underlying SN ejecta radiation contributes
significantly to the total luminosity, so we assume that CSM
interaction dominates the emergent luminosity.
%\citet{prieto17} estimated a peak bolometric luminosity of
%1.3$\times$10$^{44}$ erg s$^{-1}$ or 3.2$\times$10$^{10}$ $L_{\odot}$.
This is the peak luminosity, so we adopt $L \simeq 4 \times 10^{43}$
erg s$^{-1}$ as a rough average for $L$ over the first $\sim$100 days.
Thus, with the values $V_{\rm CSM}$ = 45 km s$^{-1}$ and $V_{\rm
  CDS}$=1600 km s$^{-1}$ adopted above, the CSM required to power the
main peak of SN~2017hcc through CSM interaction would have a wind
density parameter of $W$=2$\times$10$^{19}$ g cm$^{-1}$, corresponding
to an average mass-loss rate of 8.8$\times$10$^{25}$ g s$^{-1}$ or
$\sim$1.4 $M_{\odot}$ yr$^{-1}$.\footnote{We note that in a recent
  paper, \citet{kumar19} use a similar method to estimate a mass-loss
  rate for SN~2017hcc's progenitor, but they derive a lower value of
  0.12 $M_{\odot}$ yr$^{-1}$.  However, we note that their quoted
  luminosity of 6$\times$10$^{42}$ erg s$^{-1}$ is too low by about a
  factor of 10 for the peak absolute magnitude of $-$20.7 mag.  When
  this apparent error is corrected, their estimated $\dot{M}$ would be
  10 times larger, in good agreement with our estimate.}  Including a
bolometric correction or some efficiency factor for converting kinetic
energy to radiation would raise the CSM mass, so our estimate of the 
mass-loss rate can be considered conservative.  Of course, if the CSM 
is aspherical, then the true wind density parameter is higher than quoted 
above, but occupies only a portion of the solid angle encountered by the 
SN ejecta.

Since we do not detect significant changes in the CSM speed as the SN
evolves, this may have been a constant velocity but very short
duration wind.  The time period preceding explosion over which this
wind was active can be estimated as $t$ = $t_{\rm obs}$ $\times$
($V_{\rm CDS}/V_{\rm CSM}$).  The episodic wind must have operated for
about 6-12 years pre-explosion in order to create the CSM that powered
SN~2017hcc for the first 100-200 days.  Thus, the progenitor of
SN~2017hcc shed at least 8-16 $M_{\odot}$ in the decade before it
died.  This is comparable to estimated values of mass ejected in the
decade or so before explosion for other well-studied SLSNe like
SN~2006gy, SN~2006tf, and SN~2010jl
\citep{smith07,smith08tf,smith10gy,fransson14,woosley07}.  Based on the
late-time interaction that continued well after the main luminosity
peak, the progenitor of SN~2017hcc probably shed  mass at a
somewhat lower rate for many decades before that.

\subsection{Blueshift and Post-Shock Dust Formation}

SN~2017hcc shows the progressively increasing blueshift in its
emission-line profiles that is common in SNe~IIn.  There have been
four different suggestions for the potential origin of the systematic
blueshift seen in SNe~IIn, discussed in the next four subsections.
The only one that is a viable explanation in the case of SN~2017hcc is
the hypothesis of post-shock and/or ejecta dust formation.

\subsubsection{Radiatively accelerated CSM}

Slow pre-shock CSM could be accelerated by the tremendous
luminosity in a SLSN~IIn, and with high optical depths in the CSM, the
narrow CSM emission lines could be broadened by electron scattering to
have intermediate-width Lorentzian line profiles with a blueshifted
centroid.  This was proposed to explain the strongly blueshifted line
profiles in SN~2010jl \citep{fransson14}.  While the resulting line
shape is a symmetric Lorentzian profile, the blueshifted centroid
requires a large acceleration, and reqires that we cannot see the redshifted
CSM because it is blocked by the SN photosphere, or that the highly
asymmetric CSM is mostly on our side of the SN.  There are several problems 
in reconciling this idea with observed properties of SNe IIn:

(1) Since the Lorentzian wings originate as narrow-line photons that
are scattered and broadened by thermal electrons, the Lorentzian
profile should have the same centroid as the narrow emission, but the
observed narrow component is usually not blueshifted and remains
narrow even though the center of the Lorentzian is blueshifted.  This
is the case for both SN~2017hcc and SN~2010jl.

(2) The blueshift is small or absent at early times and becomes
progressively more pronounced at later times, but the opposite is
expected from this mechanism.  The strongest pre-shock acceleration
should occur when the SN luminosity is highest \citep{dessart15}.  In
SN~2017hcc, the narrow lines with Lorentzian wings are symmetric and
centered at zero velocity for the first $\sim$200 days as the
luminosity rises to peak and then falls.

(3) The expected amount of radiative acceleration is much smaller than
the observed shift \citep{dessart15}.  Moroever, in SN~2017hcc, the
constant narrow emission components and constant blue edges of the
P~Cyg absorption confirm very minimal (if any) pre-shock acceleration
of the CSM (less than 10 km s$^{-1}$), even though the peak of the
intermediate-width component becomes blueshifted by as much as 300-500
km s$^{-1}$, similar to SN~2010jl.

(4) The net blueshift of the intermediate-width component persists to
very late times, but as continuum optical depth drops, the ability of
electron scattering to create broad Lorentzian wings also drops.
Lines should become narrower and symmetric as the continuum optical
depth goes away, and we should see the far side of the CSM.  The
opposite is observed: the blueshift becomes progressively stronger as
the continuum fades.

(5) Electron scattering predicts no wavelength dependence, so we
should observe the same blueshift in all lines (except perhaps for
lines of different excitation levels, where as noted above, we
might expect higher speeds for higher excitation
closer to the shock).  Observations indicate, however, that the
blueshift is wavelength dependent. As noted above, the profiles of
H$\beta$, H$\alpha$, and Pa$\beta$ in SN~2017hcc indicate that the
lines become progressively more asymmetric at shorter wavelengths,
inconsistent with a cause of the blueshift being the result of
wavelength-indepenedent electron scattering.  A similar
wavelength-dependence was seen  in SN~2010jl
\citep{smith12,gall14}.

So in summary, although radiative forces may produce some
acceleration of CSM, it cannot dominate the expansion of
the CSM for reasons noted here and in Section 4.2.3.

\subsubsection{Continuum photosphere blocks redshifted side of CDS}

Occultation by the continuum photosphere of the SN could block
emission from ejecta or CSM interaction regions arising on the
redshifted side of the SN \citep{smith12}.  \citet{dessart15}
presented radiative transfer simulations that showed this effect could
produce a blueshifted emission bump in line profiles of SLSNe~IIn,
once the photosphere recedes from the pre-shock CSM and direct
emission from the post-shock CDS is revealed.  As with the previous
mechanism, however, this requires high continuum optical depths, so
this effect should be strongest at relatively early times, and the
lines should become symmetric when the photosphere recedes and the
continuum luminosity fades \citep{smith12,dessart15}.  However, in
most SNe~IIn exhibiting blueshifted lines, including SN~2017hcc and
SN~2010jl, the opposite is seen.  The blueshifted lines persist and
even become increasingly blueshifted at late times well after the
continuum luminosity has dropped by many magnitudes, as noted in the
introduction.  In SN~2017hcc, we see the most pronounced blueshift in
the intermediate-width component after day $\sim$700.  Also, because
it arises from occultation by the electron scattering photosphere,
this mechanism again predicts no wavelength dependence, clearly contradicted by observations.

\subsubsection{One-sided CSM or explosion}

In principle, a Type~IIn event might show a blueshifted
profile shape if there is stronger CSM interaction
occuring on the near side of the SN, either because of a
non-axisymmetric density distribution in the CSM (with higher
densities or smaller radii on our side), or because the explosion was
asymmetric with faster or denser SN ejecta aimed preferentially at
us.  Although one expects axisymmetric CSM from rotating
stars and binaries, one-sided CSM might not necessarily be so unusual.
Events like mergers or grazing collisions at periastron in eccentric
binaries might send a spray of CSM in one preferred direction, as 
in the outer ejecta around $\eta$~Car
\citep{kiminki16,smith18}.  Some SNe~IIn do show signs of significant
non-axisymmetric CSM and SN ejecta \citep{bilinski18}.  It is, of
course, statistically unlikely that one-sided CSM could lead SNe~IIn to show a
preference for blueshifted lines, but non-axisymmetric CSM may
nevertheless be important for individual objects. However, if the
CSM is one-sided or the explosion is lopsided,
observational clues may indicate this.

For the specific case of SN~2017hcc, one-sided CSM or a lopsided
explosion is ruled out because the narrow and broader components of
the line profiles are symmetric at early times (excluding
blushifted P Cyg absorption of course), and their continued evolution
is consistent with an axisymmetric explosion and CSM.  The
intermediate-width components of H$\beta$ and H$\alpha$ are nearly
symmetric during days 200-400, displaying only subtle blueshifted
asymmetry.  Importantly, there is no sign of asymmetry in the
Pa$\beta$ profile during this time, indicating that the intrinsic line
profile is symmetric.  But then at later times, the blueshifted
asymmetry grows even though the blue wing of the line profile stays
the same.  This behavior cannot be due to significantly one-sided CSM.
Also, even though the core of the line becomes asymmetric and
blueshifted, Figures~\ref{fig:Hafits} and \ref{fig:Halow} show that
the high-velocity wings of H$\alpha$ (beyond $\pm$3,000 km s$^{-1}$)
are symmetric.  This symmetry at high velocity indicates that there is
no significant front/back asymmetry in the fast SN ejecta, and that we
can see direct emission from the far redshifted side of the SN ejecta
(this latter point is important considering the location of the dust,
see below).  Finally, this mechanism (which depends on true asymmetry
in the gas density) predicts no systematic wavelength dependence for
the blueshift, contrary to observations.

There are certain patterns in line profile evolution that one would
expect with one-sided CSM or lopsided explosions.  For example, if
there were high-mass CSM concentrated mostly on the near side of a SN
(producing a blueshift in emission from the post-shock CDS), then we
would expect to see some corresponding and opposite asymmetry in the
broader component from the ejecta; namely, in this case the fast SN
ejecta on the blue side should hit the shock sooner and the remaining
unshocked ejecta should have slower velocities, whereas the red side
of the SN ejecta could expand less impeded, allowing us to see a
broader red wing in the SN ejecta.  This is not seen in SN~2017hcc,
although precisely this behavior was seen in SN~2012ab
\citep{bilinski18}.

Thus, while we see good evidence for axisymmetry in the CSM of
SN~2017hcc as noted earlier, there is no evidence for a one-sided CSM
density distribution or a significantly lopsided explosion that might
yield a net blueshift.

\subsubsection{Dust formation}

The formation or regrowth of dust grains is the clearly favored
explanation for the blueshift of line profiles observed in SN~2017hcc
because it self-consistently explains at least four properties that
cannot be reconciled with the previous three mechanisms: (1)
extinction by dust within the line-emitting gas is the only mechanism
consistent with the observed wavelength dependence, where H lines at
shorter wavelengths (i.e. H$\beta$ even more so than H$\alpha$) show a
stronger deficit of flux on their redshifted portions than those at
longer wavelengths (Pa$\beta$), (2) increased extinction from dust
would explain why the blue wings of the line profiles remain constant
even as their profile shapes become more blueshifted, because dust can
only absorb redshifted emission in the line but does not influence the
blue side of the line), (3) the gradual formation and buildup of dust
as the gas expands and cools is consistent with the fact that the
observed asymmetry increases with time as the SN fades, and (4) once
dust forms in the post-shock gas, it continues to cause extinction of
the redshifted line emission from receding material; this explains why
the blueshift persists to very late times, long after the continuum
luminosity has faded and the continuum optical depths have dropped.
The fact that line profiles begin symmetric and become progressively
more blueshifted with time points to dust formation within
axisymmetric material.

\begin{figure}
\includegraphics[width=2.8in]{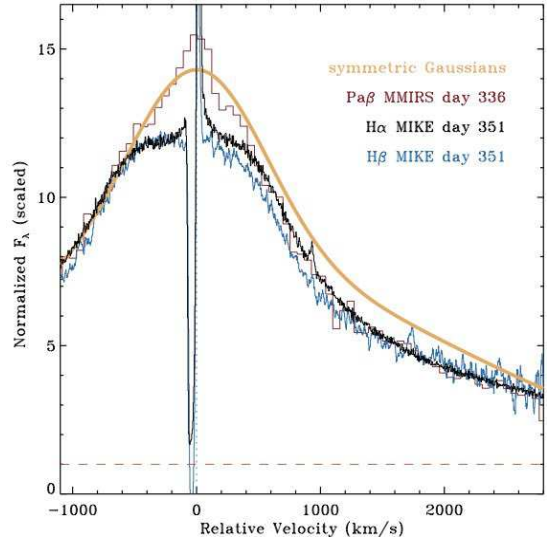}
\caption{The same day 351 H$\alpha$ and H$\beta$ profiles from
  Figure~\ref{fig:HaHb}, but including the 2-Gaussian model (orange)
  from Figure~\ref{fig:Hafits}b and the day 336 Pa$\beta$ profile (red
  histogram) from Figure~\ref{fig:Halow}.  The plotted velocity range
  zooms-in on the regions where dust influences the line profile.  The
  Gaussian model is meant to approximate the true intrinsic line
  profile shape before absorption by dust.}
\label{fig:dust}
\end{figure}

\begin{figure}
\includegraphics[width=2.8in]{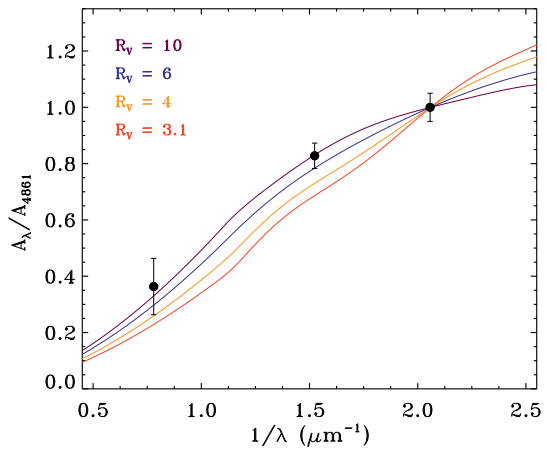}
\caption{Relative extinction derived from emission line profiles of
  Pa$\beta$, H$\alpha$, and H$\beta$, normalized to H$\beta$, compared
  to extinction laws from \citet{ccm} for various values of
  $R_V$=$A_V/E(B-V)$.}
\label{fig:extinction}
\end{figure}

SN~2017hcc's asymmetry in line profiles is admittedly more subtle than
in SN~2010jl, but the wavelength dependence is quantifiable.
Figure~\ref{fig:dust} compares the profiles of H$\beta$, H$\alpha$,
and Pa$\beta$ scaled to match the blue wings.  On the red side of the
peak, H$\beta$ is missing only about 3-4\% of the flux as compared to
H$\alpha$.  However, the difference between Pa$\beta$ and either
H$\alpha$ or H$\beta$ is more striking, with Pa$\beta$ showing only a
mild blueshift (with the caveat that the Pa$\beta$ spectrum was
obtained somewhat earlier than H$\alpha$ and H$\beta$).  We therefore
infer that the grains forming in SN~2017hcc (at least at 200-400 days)
are probably not as large and the extinction not as gray as in
SN~2010jl, where Pa$\beta$ and H$\gamma$ both showed similar
asymmetric profiles \citep{gall14}.  One could calculate the
wavelength dependence of the extinction (i.e. $R$=$A_V$/$E(B-V)$) from
the missing flux in each line, provided that the intrinsic line
profile is known.  Unlike SN~2010jl, however, SN~2017hcc was behind
the Sun when the emission from the CDS first appeared with a symmetric
profile uncorrupted by dust.  This phase was not traced in our
data, but perhaps other observations can reveal it.

If instead we take the symmetric Gaussian line profile fit in
Figure~\ref{fig:Hafits}b as a proxy for the intrinsic profile, then we
can evaluate the relative extinction in each line.  This symmetric fit
is also shown alongside the observed line profiles in
Figure~\ref{fig:dust}.  The Pa$\beta$ asymmetry at low velocities is
mild, but the broader red wing of Pa$\beta$ shows a more
significant flux deficit at +1000 to +3000 km
s$^{-1}$.  This difference may hint at larger grains that make more
gray extinction in the fast SN ejecta, and smaller grains in
the post-shock CDS.

We integrate the flux of each line between $-$1000 and +3000 km
s$^{-1}$, and taking the Gaussian fit as the intrinsic flux, we
calculate extinctions of $A_{Pa\beta}$=0.066 mag, $A_{H\alpha}$=0.148
mag, and $A_{H\beta}$=0.179 mag (note again that most of the
extinction in Pa$\beta$ occurs at the higher velocities, probably
associated with the SN ejecta and not the CDS).  These are plotted in
Figure~\ref{fig:extinction}, where the wavelength dependence of this
extinction from H emission lines is compared to a standard extinction
law \citep{ccm} with various values of $R_V$.  It is clear from
Figure~\ref{fig:extinction} that the dust in SN~2017hcc is not similar
to typical grains in the Milky Way ISM with $R_V$=3.1.  Instead, the
newly formed dust in SN~2017hcc seems to prefer larger values of $R_V$
around 6 to 10 or larger.  One could attempt a more complicated fit
with a mix of gray dust and various $R_V$ values, as demonstrated for
SN~2010jl by \citet{gall14}, but the quality of our IR data is
insufficient to provide meaningful constraints on the dust properties
at a more detailed level.  Moreover, we caution that there may be
different grain properties ($R_V$ and grain size) for the dust
forming in the post-shock CDS and in the SN ejecta, based on the
behavior of Pa$\beta$ noted above.  In any case, the data in
Figure~\ref{fig:extinction} and the large implied $R_V$ value are
sufficient to indicate the formation of rather large grains in
SN~2017hcc.  Large grains are common in the CSM of massive stars where
dust forms rapidly in eruptive events, and the formation of
relatively large grains has been inferred for other SNe IIn
\citep{gall14,nielsen18,bevan20}.  Deriving a mass of dust from the
observed extinction in lines and the grain size distribution is
complicated, as it depends on a number of assumptions about grain
properties and clumping better addressed in a sophisticated
model (e.g., \citealt{bevan18}).

But where is the new dust located?  Is dust located in the pre-shock
CSM, the SN ejecta, or the post-shock CDS?  The answer to this
question is probably: ``yes''.  There is likely to be dust in all
three zones.  Any progenitor star that has suffered enough mass loss
to make a SLSN through early CSM interaction must have experienced
very strong episodic mass loss.  Engulfed in a $\sim$10 $M_{\odot}$
cocoon, any such progenitor is likely to be surrounded by large
amounts of dust.  While such pre-existing dust in the CSM could give
rise to an IR echo when illuminated by SN
luminosity, pre-existing CSM dust cannot produce the blueshifted
asymmetry.  This requires the formation of new dust or the re-growth
behind the shock of incompletely destroyed CSM dust.

Blueshifted line profiles that become more pronounced with time could,
in principle, arise from the formation or re-growth of grains in
either the post-shock zone of the CDS or in the unshocked SN ejecta,
and both may be at work in any SN~IIn, with a differing relative
contribution to the total dust at different times.  Detailed analysis
of early to late-time optical and IR observations of SN~2010jl
demonstrated early dust formation in the post-shock CDS and continual
grain growth in the SN ejecta
\citep{smith12,gall14,sarangi18,chugai18}.

SN~2017hcc exhibits clear evidence of blueshift in the
intermediate-width component emitted by the CDS at velocities of
$-$500 to +1000 km s$^{-1}$, and it also shows a flux deficit at
faster redshifted velocites of +1000 to +3000 km s$^{-1}$.  It is
therefore likely that there is dust forming in both the CDS and
ejecta.  The blueshift of the intermediate-width component is
especially apparent at late times, and persists after the broad lines
from the SN ejecta have faded away.  Compare days 645 and 762, where
the blueshift becomes much more pronounced, even though there is
essentially no broad component left in either line.

Consider the hypothesis that dust forms only in the freely
expanding SN ejecta.  In this case, we might expect dust formation
to occur relatively late,  increasing with time.  Ejecta
dust is likely to be centrally
concentrated, as is seen in the resolved
ejecta of SN~1987A \citep{cigan19}.  In SN~2017hcc, this location
would correspond to the
orange-gradient colored inner ejecta in Figure~\ref{fig:sketch}b.
This dust could cause a blueshift in the broad component of any line
emitted by the SN ejecta, but it could also influence the
intermediate-width component by blocking some emission from redshifted
post-shock gas behind the ejecta. Thjis would remove flux mostly at the highest
redshifted speeds in the CDS.  However, it would be very unlikely for
dust in the central regions of the SN ejecta to significantlty absorb
line flux around zero velocity in the intermediate-width component,
because there is no SN ejecta along the line-of-sight to this emission
(Figure~\ref{fig:sketch}b).  In SN~2017hcc and many other interacting
SNe with these blueshifted profiles, it is evident that there is
significant missing flux in the intermediate-width component at low
velocities, often even absorbing some blueshifted velocities (shifting
the observed peak to $-$300 km s$^{-1}$ or so).  The clearest
illustration of this missing flux at low velocities is in the
difference between H$\alpha$ and Pa$\beta$, where in the IR line we
clearly see extra emission at velocities of $\pm$500 km s$^{-1}$ that
is missing in H$\alpha$ and H$\beta$.  This emission arises primarily
in portions of the CDS that are expanding near the plane of the sky,
perpendicular to our line of sight, and for which their
intermediate-width emission passes through little or no SN ejecta on
its way to us.  This blueshift cannot be due primarily to an
underlying blueshift in the broad component from the SN ejecta,
because the blueshift remains at late times (i.e. day 762 and after;
Figures~\ref{fig:Halow} and \ref{fig:Haover}) even when the broad
component has faded and only the intermediate-width H$\alpha$
component remains.  We therefore conclude that while there may be dust
forming in the SN ejecta as well, there must be some new dust
formation in the post-shock gas within the CDS of SN~2017hcc.

\section{SUMMARY AND CONCLUSIONS}

We present a series of high-resolution echelle spectra and moderate
resolution spectra of SN~2017hcc to investigate the evolution of
emission-line profiles in this highly polarized SLSN~IIn.  Here we
briefly summarize the main findings:

1.  Narrow emission and absorption in high-resolution
spectra reveal slow CSM expanding at 40-50 km.  This CSM
outflow resembles either the slow equatorial ejection in BSG binaries
(LBVs, B[e] supergiants, interacting/merging binaries, etc.) or the
dense CSM cocoons in some extreme RSGs.

2.  The narrow profiles and their variation with both time and
excitation suggest a mildly bipolar shell geometry.
Lines formed in gas at higher temperatures and ionization (He~{\sc i}
and upper Balmer series) exhibit lower sppeds, probably tracing
regions near the equator, closer to the SN
radiation and to the CSM-interaction shock in the
pinched waist.  Faster speeds (the blue edge
of the narrow P Cygni absorption) are constant, and may arise from a
polar shell. We suggest that the stark changes in narrow P Cyg
absorption strength and width arise because of the changing size and
location of the underlying radiation souerce, arising from the
relatively compact SN continuum at early times and the more distended
CSM interaction region at late times.

3.  The narrow line profiles show no evidence for significant
acceleration of the pre-shock CSM with time, and the lower velocities
seen in higher excitation/ionization lines seem to contradict
expectations for pre-shock acceleration.

4.  SN~2017hcc exhibits the typical evolution in broader emission
profiles seen in many SNe~IIn, changing from symmetric Lorentzian
profiles at early times to multi-component, asymmetric blueshifted
profiles at late times.  This reflects a change in the source of
emission from a photosphere in the pre-shock CSM at early times, when
narrow CSM lines undergo strong electron scattering that produce
symmetric profiles, to late times when the intermediate-width and
broad components of the line profile trace the post-shock CDS and SN
ejecta, respectively.

5.  Despite the high polarization reported at early times
\citep{mauerhan17}, the line profiles during the first $\sim$100 days
are remarkably symmetric, and even at later times, the observed
asymmetry in line profile shape is mild and limited to low radial
velocities. This suggests that while the explosion and CSM geometry
may be non-spherical, they are likely to be axisymmetric.

6. The time-dependence and wavelength-dependence of the blueshifted
line profiles is most likely caused by new dust formation in the
post-shock CDS and probably also the SN ejecta, with different amounts
as a function of time and possibly different grain properties in the
two regions.  For various reasons discussed in detail, the blueshift
cannot be due to pre-shock acceleration of CSM, the SN continuum
blocking far side of the CDS, or one-sided CSM/ejecta.
Instead, the blueshift must be due to dust formation.  The degree of
blueshifted asymmetry has a wavelength dependence consistent with
extinction from dust, with a stronger effect in lines at shorter
wavelengths.  The wavelength-dependent extinction suggests rather
large grains, consistent with a total-to-selective extinction ratio of
$R_V$=6-10 or more.

7.  From the observed expansion speeds of the CSM and CDS, combined with published estimates of the luminosity, we infer a
high mass-loss rate of roughly 1 $M_{\odot}$ yr$^{-1}$ in the decade
or so before explosion, with a CSM shell of around 10 $M_{\odot}$ or
more.  This is quite similar to extreme CSM mass values derived from
a number of other SLSNe~IIn.  There may have been strong mass loss for
decades or centuries before that as well, judging from the ongoing
CSM interaction and constant P Cyg absorption velocity from the CSM.

While we have shown that the blueshifted emission-line profiles in
SN~2017hcc must arise from the formation of relatively large dust
grains in the CDS and SN ejecta, we expect that additional
observations (especially in the IR) can improve our understanding of
SN~2017hcc.  Namely, better IR spectra over more epochs and multi-band
IR photometry, combined with modeling of the line profiles, could
significantly improve detailed estimates for the grain properties as
well as the relative amounts of dust in the pre-existing CSM, the CDS,
and the SN ejecta.  Reconciling the nearly symmetric line profiles of
SN~2017hcc (which imply mild asymmetry in the form of axisymmetric
CSM) with its very high polarization may be an interesting challenge
that leads to new insights about pre-SN mass loss in SLSNe~IIn.

\section*{Acknowledgements}
%\scriptsize
We thank an anonymous referee for helpful suggestions.  
Support for NS was provided by NSF award AST-1515559, and by the
National Aeronautics and Space Administration (NASA) through HST grant
AR-14316 from the Space Telescope Science Institute, operated
by AURA, Inc., under NASA contract NAS5-26555. Some data reported 
here were obtained at the MMT Observatory, a joint facility of the 
University of Arizona and the Smithsonian Institution. This paper 
includes data gathered with the 6.5 meter Magellan Telescopes located at 
Las Campanas Observatory, Chile.

Facilities:  MMT (Bluechannel, Redchannel, Binospec, MMIRS), Magellan (MIKE, IMACS), Bok (B\&C)

\section*{Data availability} 
The data underlying this article will be shared on reasonable request to the corresponding author.

\end{document}